\documentclass[aps,prd,twocolumn,superscriptaddress,nofootinbib,floatfix,showkeys]{revtex4-2}

\usepackage[T1]{fontenc}
\usepackage{bm}
\usepackage{graphicx}
\usepackage{amsmath,amssymb,amsfonts}
\usepackage{multirow}
\usepackage{array}
\usepackage{etoolbox}
\usepackage{orcidlink}
\usepackage{xcolor}
\usepackage{hyperref}
\hypersetup{colorlinks=true,linkcolor=red,citecolor=blue,urlcolor=blue}

\begin{document}
\title{Quasinormal Modes and Environmental Trapping around Charged Letelier Black Holes with Perfect-Fluid Dark Matter and Quintessence}

\author{Semra Gurtas Dogan\orcidlink{0000-0001-7345-3287}}
\email{semragurtasdogan@hakkari.edu.tr (Corresponding Author)}
\affiliation{Department of Medical Imaging Techniques, Hakkari University, 30000 Hakkari, T\"urkiye}

\author{Abdullah Guvendi\orcidlink{0000-0003-0564-9899}}
\email{abdullah.guvendi@erzurum.edu.tr}
\affiliation{Department of Basic Sciences, Erzurum Technical University, 25050 Erzurum, T\"urkiye}

\author{Hassan Hassanabadi\orcidlink{0000-0001-7487-6898}}
\email{hha1349@gmail.com}
\affiliation{Department of Physics, Faculty of Science, University of Hradec Králové, Rokitanského 62, 500 03 Hradec Králové, Czechia}
\affiliation{Physics Department, California State University, Fresno, CA 93740, USA}

\begin{abstract}
We compute the massless-scalar quasinormal spectra of charged Letelier black holes surrounded by a logarithmic perfect-fluid-dark-matter (PFDM) component and quintessence in both the $\Lambda =0$ and asymptotically anti-de Sitter (AdS) ($\Lambda <0$) regimes, together with the test-Maxwell spectra in the $\Lambda =0$ regime. For $\Lambda =0$, sixth-order WKB calculations with Pad\'e resummation are validated against independent time-domain evolutions, yielding agreement within $0.06\%$ for the fundamental modes examined. The parameter scans show that increasing the string-cloud and quintessence strengths decreases both the oscillation frequencies and the damping rates, whereas increasing the electric charge or the positive PFDM parameter increases both quantities. In the adopted convention, positive $\beta$ corresponds to negative PFDM energy density, whereas negative $\beta$ gives positive energy density and reverses the frequency shift. The angular frequency and Lyapunov exponent of the unstable circular null orbit reproduce the leading eikonal trends and approximate the $\ell=2$ WKB frequencies at the percent level. For $\Lambda <0$, where the scalar effective potential diverges at the AdS boundary and the standard single-barrier WKB prescription is inapplicable, we solve the spectral problem using a Chebyshev--Gauss--Lobatto pseudospectral method in ingoing Eddington--Finkelstein coordinates, imposing regularity at the future horizon and the normalizable (Dirichlet) boundary condition. The resulting scalar modes are independently checked against time-domain evolutions and established Schwarzschild--AdS benchmarks. The AdS spectrum contains long-lived modes trapped between the photon-sphere barrier and the reflecting boundary, with damping rates spanning approximately five orders of magnitude. Their damping correlates with the ratio of the oscillation frequency to the barrier-top frequency, supporting an interpretation in terms of leakage through the potential barrier. No growing modes are found in the test-field sectors and parameter ranges investigated; this does not establish the stability of the fully coupled gravitational--electromagnetic--matter system.
\end{abstract}


\maketitle

\section{Introduction}
\label{sec:intro}

Quasinormal modes (QNMs) are the characteristic damped oscillations of a perturbed black hole. Because their frequencies are fixed by the black hole and by whatever surrounds it, they are a natural tool to probe the strong field region, and they dominate the ringdown of the gravitational wave signals observed since 2015~\cite{LIGO2016}. The theory of QNMs and the methods used to compute them are reviewed in Refs.~\cite{Kokkotas1999,Berti2009,Konoplya2011}.

A black hole in nature is not isolated, and three simple ingredients are often used to model its environment. The first is a cloud of strings, introduced by Letelier~\cite{Letelier1979}, which adds a constant term to the lapse function and produces a solid angle deficit. The second is quintessence, modelled as an anisotropic fluid following Kiselev~\cite{Kiselev2003}, which contributes a radial term controlled by its normalization $\sigma$ and equation-of-state parameter $\omega$, with $-1<\omega <-1/3$. The third is perfect fluid dark matter (PFDM)~\cite{LiYang2012,Kiselev2003b,Xu2018}, whose contribution is logarithmic and is controlled by the PFDM parameter $\beta$. We combine these ingredients with an electric charge $q$ and a cosmological constant $\Lambda$. For the sources considered here, the corresponding contributions to the mass function add linearly~\cite{Kiselev2003}, so the lapse function is obtained by summing the individual terms. This charged Letelier--AdS geometry with PFDM and quintessence was obtained in Ref.~\cite{EslamPanah2026}, where its horizon thermodynamics, ray optics and scalar wave propagation were studied. For the scalar field that work identified frequency dependent barriers and candidate trapping regions between the photon sphere barrier and the AdS boundary, and pointed out that the complex quasinormal frequencies require a separate boundary value calculation.

Several special cases of this spacetime have been studied before. QNMs of a Schwarzschild black hole with a cloud of strings in a PFDM background~\cite{Hamil2025}, and of a Schwarzschild black hole in PFDM with quintessence~\cite{Hamil2025b}, were computed with the WKB method. Scalar QNMs of a charged black hole in PFDM were obtained in Ref.~\cite{Das2024}, together with their eikonal limit. The QNMs of a charged AdS black hole in PFDM, without strings and quintessence, were obtained with a Chebyshev pseudospectral method and related to its thermodynamic phase transitions in Ref.~\cite{LiFeng2026,BertiKokkotas2003}. The uncharged Letelier--AdS black hole in PFDM with quintessence was considered in Ref.~\cite{Ahmed2025}, where geodesics, shadows and thermodynamics, but not QNMs, were discussed. Thermodynamics, shadows and emission rates were also studied for a charged AdS black hole with a string cloud and quintessence~\cite{Toledo2019,Wang2026} and for an Euler--Heisenberg AdS black hole with a string cloud, PFDM and quintessence~\cite{AhmedSilva2026}, whose lapse function reduces to the one studied here when the Euler--Heisenberg coupling vanishes. To the best of our knowledge, the quasinormal-mode spectrum of a charged black hole incorporating all of these matter contributions, the geometry of Ref.~\cite{EslamPanah2026}, has not yet been systematically investigated. The present work addresses this gap by providing a comprehensive analysis of the corresponding perturbation spectra.

The physical interpretation of the matter sector in this spacetime requires careful consideration. Because the temporal and radial metric components satisfy ${g}_{tt}{g}_{rr}=-1$, each source has radial pressure equal to minus its energy density. For the positive sign of the PFDM parameter adopted in the principal parameter scans, the PFDM contribution has negative energy density, as also noted in Ref.~\cite{EslamPanah2026}. The sources are therefore effective anisotropic fluids rather than halos of collisionless particles, and we treat the spacetime as a phenomenological model. To see how much the predicted shifts of the spectrum depend on this choice, we also compute the fundamental modes for negative values of the PFDM parameter, for which its energy density is positive.

We consider two regimes. For $\Lambda =0$ and nonzero quintessence normalization with $\omega =-2/3$, the quintessence term produces a cosmological horizon; when $\sigma =0$, the exterior instead extends to spatial infinity. This is also the setting of the related studies mentioned above~\cite{Hamil2025,Hamil2025b,Das2024}. For the cosmological-horizon configurations, the effective potential is a single barrier between the event and cosmological horizons, where the WKB method is well suited. We compute the scalar and electromagnetic frequencies with the sixth order WKB formula and its Pad\'e improved form~\cite{Konoplya2003,Matyjasek2017,Konoplya2019} and check them with time-domain integration~\cite{Gundlach1994} and the matrix pencil method~\cite{Hua1990}. For $\Lambda <0$ the scalar potential diverges at the AdS boundary, so the outgoing-wave assumptions underlying the single-barrier WKB formula do not apply. We instead solve the boundary-value problem with a Chebyshev spectral method~\cite{Trefethen2000,Jansen2017}, using the standard AdS quasinormal-mode prescription and the Schwarzschild--AdS benchmarks of Horowitz and Hubeny~\cite{Horowitz2000}, together with later charged and uncharged AdS studies~\cite{BertiKokkotas2003,Batic2026}; an independent time-domain evolution provides an additional check.

This paper is structured as follows. We first define the geometry and its effective matter sources in Sec.~\ref{sec:geometry}, then derive the scalar and test-Maxwell perturbation equations in Sec.~\ref{sec:pert}. The cosmological-horizon problem is treated with WKB and time-domain methods in Secs.~\ref{sec:wkb} and~\ref{sec:td}; the AdS eigenvalue problem and its cavity interpretation are developed in Secs.~\ref{sec:ads} and~\ref{sec:cavity}. Section~\ref{sec:conc} brings these results together and states the scope and limitations of the analysis. We use $G=c=1$ and set ${m}_{0}=1$.

\section{The spacetime}
\label{sec:geometry}

We consider a static, spherically symmetric geometry containing a cloud of strings, a quintessence component, a logarithmic perfect-fluid-dark-matter (PFDM) contribution, an electromagnetic field, and a cosmological constant. The uncharged Letelier--AdS geometry with logarithmic PFDM and quintessence was studied by Ahmed and Silva~\cite{Ahmed2025}. Its string-cloud, quintessence, and PFDM terms are rooted, respectively, in the source constructions of Letelier, Kiselev, and Li and Yang~\cite{Letelier1979,Kiselev2003,LiYang2012}. Its charged extension, with the Maxwell term ${q}^{2}/{r}^{2}$ familiar from the Reissner--Nordstr\"om family and its AdS counterpart~\cite{BertiKokkotas2003}, was obtained in Ref.~\cite{EslamPanah2026}. The resulting lapse function and the associated effective source decomposition are specified below.

The line element is
\begin{equation}
ds^2=-f(r)\,dt^2+\frac{dr^2}{f(r)}+r^2\left(d\theta^2+\sin^2\theta\,d\varphi^2\right),
\label{eq:metric}
\end{equation}
with the lapse function~\cite{EslamPanah2026}
\begin{equation}
f(r)=1-\alpha-\frac{2m_0}{r}-\frac{\Lambda r^2}{3}+\frac{q^2}{r^2}+\frac{\beta}{r}\ln\frac{r}{|\beta|}-\frac{\sigma}{r^{3\omega+1}} .
\label{eq:f}
\end{equation}
Here ${m}_{0}$ is the mass parameter, $q$ the electric charge, $\alpha$ the string-cloud parameter, $\beta$ the PFDM parameter, and $\sigma$ and $\omega$ the normalization and equation-of-state parameter of the quintessence field. For $q=0$, Eq.~\eqref{eq:f} reduces to the uncharged Letelier--AdS--PFDM--quintessence solution of Ref.~\cite{Ahmed2025}; for $\alpha =\beta =\sigma =0$, it reduces to the Reissner--Nordstr\"om--AdS metric. The term $-\alpha$ is the Letelier string-cloud contribution~\cite{Letelier1979}; the term $-\sigma {r}^{-(3\omega +1)}$ is the Kiselev quintessence contribution~\cite{Kiselev2003}; and the logarithmic term is the PFDM contribution in the convention of Li and Yang~\cite{LiYang2012}. The Maxwell field supplies ${q}^{2}/{r}^{2}$, as in Reissner--Nordstr\"om--AdS geometries~\cite{BertiKokkotas2003}. The uncharged limit therefore recovers the geometry of Ahmed and Silva~\cite{Ahmed2025}, while the source-by-source decomposition of the charged solution follows from the Einstein equations and is given in Sec.~\ref{sec:matter} (see also Ref.~\cite{EslamPanah2026}). Equation~\eqref{eq:f} is also the limit of vanishing Euler--Heisenberg coupling of the solution of Ref.~\cite{AhmedSilva2026}, with their PFDM parameter $\lambda$, quintessence normalization $N$ and AdS radius $\ell_p$ related to ours by $\lambda=\beta$, $N=\sigma$ and $1/\ell_p^2=-\Lambda/3$. For $\beta=0$ it reduces to the charged AdS black hole with a cloud of strings and quintessence of Ref.~\cite{Toledo2019}.

\begin{figure}[tbp]
\centering
\includegraphics[width=\columnwidth]{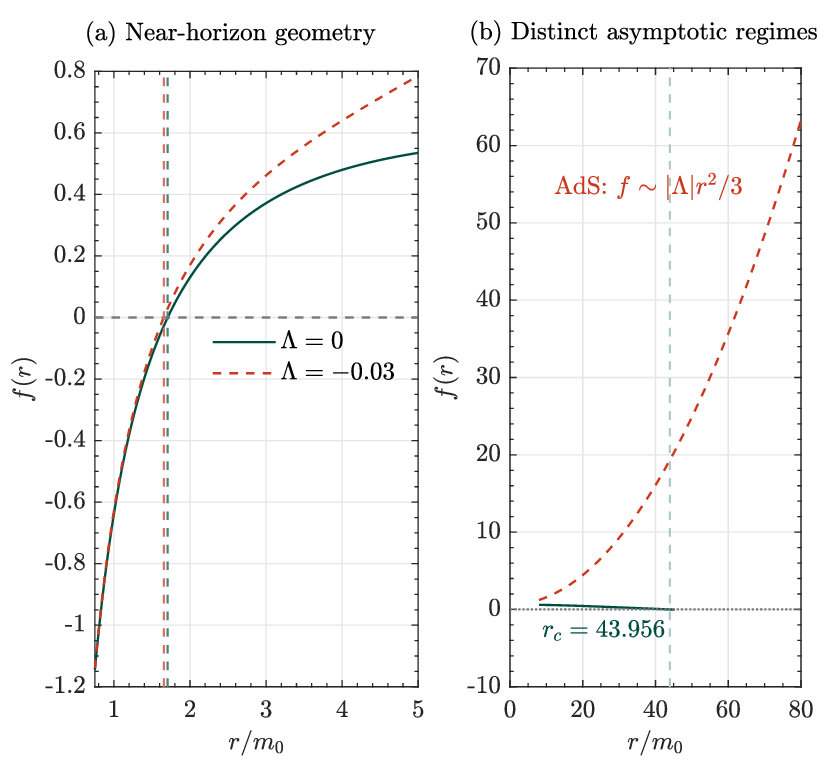}
\caption{Radial lapse function $f(r)$ in Eq.~\eqref{eq:f} for the reference parameters in Eq.~\eqref{eq:ref}, with ${m}_{0}=1$, $\alpha =0.1$, $\beta =0.2$, $q=0.4$, $\sigma =0.02$, and $\omega =-2/3$. Panel (a) resolves the near-horizon region and compares the $\Lambda =0$ geometry with the AdS reference geometry ($\Lambda =-0.03$); the vertical markers identify the corresponding event horizons ${r}_{+}=1.70634028$ and ${r}_{+}=1.65563890$. Panel (b) displays the distinct large-radius asymptotics: for $\Lambda =0$ the quintessence term produces the cosmological horizon ${r}_{c}=43.95596099$, whereas for $\Lambda <0$ the lapse grows as $f(r)\sim |\Lambda |{r}^{2}/3$. The three positive roots in the $\Lambda =0$ case are ${r}_{-}=0.07518356$, ${r}_{+}=1.70634028$, and ${r}_{c}=43.95596099$.}
\label{fig:lapse}
\end{figure}

We take the reference values of Ref.~\cite{EslamPanah2026},
\begin{equation}
\begin{aligned}
m_0&=1, & \alpha&=0.1, & \beta&=0.2, & q&=0.4,\\
\sigma&=0.02, & \omega&=-\tfrac23, & \Lambda&=-0.03.
\end{aligned}
\label{eq:ref}
\end{equation}
and vary one parameter at a time: $\sigma \in \{0,0.06,0.10\}$, $\alpha \in \{0,0.2,0.3\}$, $q\in \{0,0.6,0.8\}$, $\beta \in \{0.05,0.4,0.6\}$ and $\omega \in \{-0.8,-0.5\}$. In the AdS regime we also consider $\Lambda =-0.01$ and $\Lambda =-0.05$. For the reference set the event horizon lies at ${r}_{+}=1.6556$, in agreement with Ref.~\cite{EslamPanah2026}. The second regime is obtained by setting $\Lambda =0$ and keeping the other parameters. The lapse function for both cases is plotted in Fig.~\ref{fig:lapse}.

\subsection{Units and choice of parameters}

All quantities are measured in units of ${m}_{0}$. Under the rescaling $r\rightarrow {m}_{0}r$, $t\rightarrow {m}_{0}t$ the lapse function keeps its form, with $\alpha$, $q/{m}_{0}$, $\beta /{m}_{0}$, $\sigma {m}_{0}^{3\omega +1}$ and $\Lambda {m}_{0}^{2}$ as the dimensionless parameters, so that ${m}_{0}\varpi$ depends only on these combinations. In physical units the oscillation frequency is
\begin{equation}
\nu=\frac{{\rm Re}(m_0\varpi)}{2\pi}\,\frac{c^3}{Gm_0}\simeq32.3\,{\rm kHz}\times{\rm Re}(m_0\varpi)\times\frac{M_\odot}{m_0}.
\label{eq:nu}
\end{equation}
The frequency $\varpi$ refers to the time coordinate $t$, and a static observer at radius ${r}_{o}$ measures $\varpi /\sqrt{f({r}_{o})}$. For $\sigma =\Lambda =0$, $f\rightarrow 1-\alpha$ at large $r$ and a distant observer measures $\varpi /\sqrt{1-\alpha }$, as in other spacetimes with a solid angle deficit~\cite{Barriola1989}. With a cosmological horizon there is no asymptotic region, so Eq.~\eqref{eq:nu} is only an order of magnitude estimate. The fundamental $\ell=2$ scalar mode of the reference set with $\Lambda =0$ (Sec.~\ref{sec:wkb}) then corresponds to about $1.7$\,kHz for ${m}_{0}=10{M}_{\odot }$ and to about $4$\,mHz for ${m}_{0}=4\times 10^{6}{M}_{\odot }$.

These parameter values  of Ref.~\cite{EslamPanah2026} are large, so that the effect of each matter field is easy to see. They are not estimates for astrophysical black holes, whose environment is expected to be much more dilute. The frequencies vary smoothly with all parameters (Sec.~\ref{sec:wkb}), and for small values the variation is close to linear. For example, for $\Lambda =0$, going from $\sigma =0$ to $\sigma =0.02$ lowers ${\rm Re}\,\varpi$ and $|{\rm Im}\,\varpi|$ of the fundamental $\ell=2$ scalar mode by $8\%$ and $10\%$, so a quintessence normalization ten times smaller would change them by roughly $1\%$. For smaller parameters the shifts can therefore be estimated by linear scaling.

Near the black hole the two curves are almost identical, but their behaviour at large $r$ is completely different. With $\Lambda =-0.03$ the term $-\Lambda {r}^{2}/3$ dominates and $f$ grows without bound. With $\Lambda =0$ and $\omega =-2/3$ the quintessence term is $-\sigma r$, so that $f\simeq 1-\alpha -\sigma r$ at large distances and a cosmological horizon appears at ${r}_{c}\approx (1-\alpha )/\sigma$. For the reference values we find three horizons: an inner one at ${r}_{-}=0.0752$, the event horizon at ${r}_{+}=1.7063$ and the cosmological horizon at ${r}_{c}=43.956$. The static region ${r}_{+}<r<{r}_{c}$ is where the perturbations live. The surface gravity of the cosmological horizon, ${\kappa }_{c}=|f'({r}_{c})|/2=0.0097$, will be needed in Sec.~\ref{sec:td}. For $\sigma =0$ there is no cosmological horizon and $f\rightarrow 1-\alpha$ at infinity, up to the slowly decaying PFDM term. For $\omega =-0.8$ and $\omega =-0.5$ the quintessence term grows as ${r}^{1.4}$ and ${r}^{0.5}$, and the cosmological horizon moves to ${r}_{c}=14.19$ and ${r}_{c}=2024.7$, respectively. The horizon radii are evaluated for each parameter set and are reported with the photon-sphere data below.

\subsection{Matter content}
\label{sec:matter}

The metric function alone does not determine whether the environmental terms represent physically admissible matter. We therefore reconstruct the effective stress tensor and examine the associated energy density, principal pressures, and energy conditions. This step also fixes the sign interpretation of the logarithmic PFDM contribution used in the spectral analysis. To see what kind of matter the lapse function~\eqref{eq:f} describes, we write $f=1-2m(r)/r$, with
\begin{equation}
m(r)=m_0+\frac{\alpha r}{2}-\frac{q^2}{2r}-\frac{\beta}{2}\ln\frac{r}{|\beta|}+\frac{\sigma}{2}\,r^{-3\omega}+\frac{\Lambda r^3}{6}.
\label{eq:mass}
\end{equation}
For a metric of the form~\eqref{eq:metric} the field equations give ${T}^{t}{}_{t}={T}^{r}{}_{r}$, so every source has energy density $\rho =m'/(4\pi {r}^{2})$, radial pressure ${P}_{r}=-\rho$ and tangential pressure ${P}_{t}=-\rho -r\rho '/2$~\cite{Visser2020}. The Maxwell field gives $\rho ={q}^{2}/(8\pi {r}^{4})$ and ${P}_{t}=\rho$, the string cloud $\rho =\alpha /(8\pi {r}^{2})$ and ${P}_{t}=0$~\cite{Letelier1979,Ahmed2025}, and the quintessence term $\rho =-3\omega \sigma /(8\pi {r}^{3(1+\omega )})$ and ${P}_{t}=(3\omega +1)\rho /2$~\cite{Kiselev2003}. For $\alpha >0$, $\sigma >0$ and $-1<\omega <-1/3$ these densities are positive, and the null, weak and dominant energy conditions hold; the quintessence term violates the strong energy condition, $\rho +{P}_{r}+2{P}_{t}=(3\omega +1)\rho <0$, as expected for a source of cosmic acceleration. As pointed out by Visser~\cite{Visser2020}, the Kiselev source is an anisotropic fluid and not a perfect fluid, and $\omega$ is the ratio of the averaged pressure to the energy density.

The PFDM term gives
\begin{equation}
\rho=-\frac{\beta}{8\pi r^3},\qquad P_r=-\rho,\qquad P_t=\frac{\rho}{2}.
\label{eq:pfdm}
\end{equation}
Comparison with the Maxwell term, which enters $m(r)$ in the same way and has a positive energy density, fixes this sign independently of the convention for the field equations, so the PFDM energy density is positive only for $\beta <0$. In Ref.~\cite{Xu2018} the energy density of the same term was given with the opposite sign, $+\alpha /(8\pi {r}^{3})$ for $a=0$, with the field equations written as ${G}_{\mu \nu }=-8\pi {T}_{\mu \nu }$. The positive sign of the logarithmic term, used  in Ref.~\cite{EslamPanah2026} and here, is the one for which Xu \emph{et al.} found that the rotation velocity of circular orbits tends to a constant~\cite{Xu2018}, the property that originally motivated the PFDM term~\cite{LiYang2012,Kiselev2003b}. For $\beta >0$, however, $\rho <0$ and $\rho +{P}_{t}=3\rho /2<0$, so the weak and null energy conditions are violated, consistent with the origin of the logarithmic term in a phantom field~\cite{LiYang2012}. The PFDM stress tensor is also anisotropic, ${P}_{r}\ne {P}_{t}$. For the black hole of Ref.~\cite{Xu2018} with $\Lambda =0$, horizons exist for $0<\beta <2{m}_{0}$ and for $-7.18{m}_{0}<\beta <0$; all values of $\beta$ used here lie within these ranges. For $r>|\beta |$ the PFDM term with $\beta >0$ lowers $m(r)$; at the photon sphere of the reference set it contributes $-0.254$ to $m({r}_{ph})$.

Since ${T}^{t}{}_{t}={T}^{r}{}_{r}$ for every source, the stress tensor in the $(t,r)$ plane is proportional to the metric, and every radially moving observer measures the same energy density. The sources therefore do not single out a static rest frame, and, like the electric field of the Reissner--Nordstr\"om solution, they extend smoothly inside the horizon. None of the matter terms has a finite total mass. Even for $\sigma =\Lambda =0$, $m(r)$ grows as $\alpha r/2$ and contains a logarithm, so ${m}_{0}$ is a parameter of the solution rather than an ADM mass.

Since only the total stress tensor enters the field equations, we also checked the energy conditions for the sum of all terms in the static region ${r}_{+}<r<{r}_{c}$, for every parameter set with $\Lambda =0$. For the reference set the total energy density and $\rho +{P}_{t}$ stay positive, because the string cloud, the Maxwell field and the quintessence term outweigh the negative PFDM density, so the null and weak energy conditions hold for the total source although they fail for the PFDM term alone. The same is true for all remaining parameter sets except $\alpha =0$, $q=0$, $\beta =0.4$ and $\beta =0.6$. In these four cases the PFDM term dominates in part of the static region and $\rho +{P}_{t}$ becomes negative; for $\alpha =0$, $\beta =0.4$ and $\beta =0.6$ the total energy density is negative there as well. For $\beta <0$ every term satisfies both conditions on its own. For $\Lambda <0$ the cosmological constant adds the constant density $\Lambda /(8\pi )<0$ with $\rho +{P}_{t}=0$. This leaves the null energy condition unchanged, but for the reference AdS set it makes the total energy density negative for $r>2.0$, as expected for a negative cosmological constant.

\section{Perturbation equations}
\label{sec:pert}

Having established the background and its matter interpretation, we now formulate the linear test-field problem. The scalar and electromagnetic sectors are treated separately on the prescribed geometry; this distinction is particularly important for a charged background, where electromagnetic and gravitational perturbations can couple in the fully dynamical system.

A massless scalar field obeys ${\partial }_{\mu }(\sqrt{-g}{g}^{\mu \nu }{\partial }_{\nu }\Phi )=0$. With $\Phi=e^{-i\varpi t}\,Y_{\ell m}(\theta,\varphi)\,\psi(r)/r$ and the tortoise coordinate $d{r}_{*}=dr/f$, one finds
\begin{equation}
\frac{d^2\psi}{dr_*^2}+\left[\varpi^2-V(r)\right]\psi=0 .
\label{eq:schr}
\end{equation}
We denote the frequency by $\varpi$ to avoid confusion with the quintessence parameter $\omega$. The electromagnetic field on a background with ${g}_{tt}{g}_{rr}=-1$ leads to the same equation~\cite{Konoplya2011}, and the two potentials can be written together as
\begin{equation}
V(r)=f(r)\left[\frac{\ell(\ell+1)}{r^2}+\left(1-s^2\right)\frac{f'(r)}{r}\right],
\label{eq:V}
\end{equation}
with $s=0$ for the scalar field and $s=1$ for the electromagnetic field. The electromagnetic field is treated as a test field. In a charged background electromagnetic and gravitational perturbations mix, and for the Reissner--Nordstr\"om black hole the coupled problem has been solved~\cite{Leaver1990}. To estimate the effect we computed, in the Reissner--Nordstr\"om limit ($\alpha =\beta =\sigma =\Lambda =0$), the electromagnetic type modes from the axial potentials of Ref.~\cite{Leaver1990} with the same WKB code; they reproduce the values of Ref.~\cite{Leaver1990} to within $5\times 10^{-4}$. For $q=0.4$ the coupling raises ${\rm Re}\,\varpi$ of the fundamental mode by $2.2\%$ ($\ell=1$) and $1.9\%$ ($\ell=2$), and $|{\rm Im}\,\varpi|$ by less than $1\%$; for $q=0.8$ the shifts of the real part grow to $12\%$ and $8\%$. Our electromagnetic results are therefore those of a test Maxwell field, and for a coupled field they are accurate to a few per cent only at small $q$. The coupled problem with strings, PFDM and quintessence would require perturbing the matter sources as well. The scalar results are not affected.

\begin{figure}[tbp]
\centering
\includegraphics[width=0.85\columnwidth]{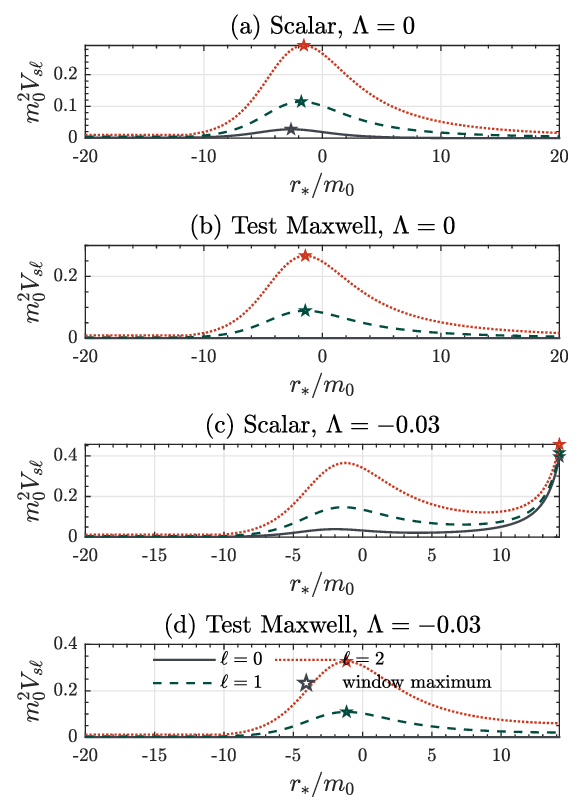}
\caption{Effective potentials $m_0^2V_{s\ell}$ from Eq.~\eqref{eq:V} as functions of the dimensionless tortoise coordinate ${r}_{*}/{m}_{0}$, with the origin fixed by ${r}_{*}=0$ at $r=3{m}_{0}$. The four panels compare (a) scalar perturbations ($s=0$) for $\Lambda =0$, (b) test-Maxwell perturbations ($s=1$) for $\Lambda =0$, (c) scalar perturbations for the AdS reference geometry ($\Lambda =-0.03$), and (d) test-Maxwell perturbations for the same AdS background. Each panel shows $\ell=0,1,2$; line styles distinguish the multipoles and the star symbols mark the maximum within the displayed radial/tortoise-coordinate window. The accompanying labels report the corresponding numerical $V$ value and ${r}_{*}$ position. For the scalar AdS potential the boundary behavior is divergent, so the marked endpoint-window maximum is not a finite global barrier maximum and is not the $\varpi_{\rm top}$ used in the cavity analysis. The $\Lambda =0$ potentials vanish at both horizons, whereas the AdS scalar potential rises toward the reflecting boundary. The background parameters are those of Eq.~\eqref{eq:ref}, with $\Lambda$ set as indicated in each panel.}
\label{fig:pot}
\end{figure}

The shape of the potential is shown in Fig.~\ref{fig:pot}.

 For $\Lambda =0$ the potential vanishes at both horizons (${r}_{*}\rightarrow \pm \infty$) and has a single maximum close to the photon sphere. The QNMs are defined by the boundary conditions
\begin{align}
\psi&\sim e^{-i\varpi r_*}\quad (r_*\to-\infty),\\
\psi&\sim e^{+i\varpi r_*}\quad (r_*\to+\infty),
\label{eq:bc0}
\end{align}
i.e.\ purely ingoing at the event horizon and purely outgoing at the cosmological horizon. This is the standard setting for the WKB method.

For $\Lambda <0$ the situation is different. The tortoise coordinate now tends to a finite value ${r}_{*}^{\infty }$ as $r\rightarrow \infty$ (${r}_{*}^{\infty }=16.44$ for the reference set with ${r}_{*}=0$ at $r=3$), and for the scalar field $V\rightarrow +\infty$ there, since $f\simeq |\Lambda |{r}^{2}/3$ and $V\simeq ff'/r\propto {r}^{2}$. After the barrier near the photon sphere the potential goes through a shallow minimum and then rises to form a reflecting wall. The QNMs are now the solutions that are ingoing at the horizon and vanish at the AdS boundary~\cite{Horowitz2000}. In this regime we restrict ourselves to the scalar field. For the electromagnetic field the potential tends to a finite constant at the boundary, so the choice of boundary condition there requires a separate discussion; requiring, for instance, a vanishing energy flux at the boundary leads to two different reflective branches~\cite{Batic2026,Wang2015}. We do not attempt this here. The WKB formula of Sec.~\ref{sec:wkb} matches solutions that are purely outgoing on both sides of a single barrier; at the AdS boundary the wave is fully reflected, so a different method is needed (Sec.~\ref{sec:ads}).

\section{WKB spectrum for \texorpdfstring{$\Lambda =0$}{Lambda = 0}}
\label{sec:wkb}

We begin with the cosmological-horizon case, for which the effective potential is a localized barrier and the quasinormal boundary conditions are outgoing at the two asymptotic ends of the tortoise coordinate. This structure permits a systematic WKB treatment. We use the sixth-order approximation and its Pad\'e improvement, and assess their reliability through independent checks described below.

\subsection{Method}

The WKB construction approximates the potential in the neighbourhood of its maximum and expresses the resonance condition in terms of its derivatives there. We state the quantization condition first, then describe how the required derivatives and truncation uncertainties are evaluated for the present non-polynomial lapse function.

For a single barrier the QNMs follow from the WKB condition~\cite{Konoplya2003,Schutz1985,Iyer1987}
\begin{equation}
\frac{i\left(\varpi^2-V_0\right)}{\sqrt{-2V_0''}}-\sum_{k=2}^{6}\Lambda_k=n+\frac12 ,
\label{eq:wkb}
\end{equation}
where ${V}_{0}$ and ${V}_{0}''$ are the value and the second ${r}_{*}$-derivative of the potential at its maximum and the corrections ${\Lambda }_{k}$ depend on higher derivatives. Since the sixth order formula needs derivatives of $V$ up to order twelve, we located the maximum ${r}_{0}$ numerically, expanded $f$ in a Taylor series about ${r}_{0}$ to high order, and applied the WKB routines of Ref.~\cite{Konoplya2019} to the resulting polynomial in $x=r-{r}_{0}$. This also handles the logarithmic PFDM term and the noninteger powers that appear for $\omega \ne -2/3$. The same sixth order frequencies were also obtained with an independent code that solves the WKB recursion order by order from the Taylor coefficients of $V({r}_{*})$; for all 120 computed entries the two codes agree to better than $10^{-6}$. In the Schwarzschild limit ($\alpha =\beta =q=\sigma =0$) we obtain $0.292910-0.097762i$ and $0.483642-0.096766i$ for the scalar $\ell=1$ and $\ell=2$ fundamental modes, the known sixth order values~\cite{Konoplya2003}.

As an estimate of the WKB error we use $\Delta_6=\tfrac12\left(|\varpi_7-\varpi_6|+|\varpi_6-\varpi_5|\right)$, where ${\varpi }_{k}$ is the $k$th order result~\cite{Konoplya2019}. For the fundamental modes ${\Delta }_{6}$ is below $4\times 10^{-4}$ for the scalar and $7\times 10^{-4}$ for the electromagnetic $\ell=1$ modes, and below $5\times 10^{-5}$ for $\ell=2$. The least accurate entry is the first overtone of the electromagnetic $\ell=1$ mode, with ${\Delta }_{6}$ up to $1.3\times 10^{-3}$. We also computed the Pad\'e averaged values of the same WKB series~\cite{Matyjasek2017,Konoplya2019}, which usually improve the accuracy. For the fundamental modes they differ from the sixth order values by at most $5.3\times 10^{-4}$ for $\ell=1$ and $3\times 10^{-5}$ for $\ell=2$, consistent with ${\Delta }_{6}$. Both values are listed in the tables.

\subsection{Results}

We now collect the scalar and electromagnetic frequencies and compare their response to variations of the background parameters. The tables report both the direct sixth-order WKB values and the Pad\'e-averaged values, allowing the numerical size of the resummation correction to be assessed alongside the physical parameter trends.

\begin{table*}[tbp]\centering\footnotesize\setlength{\tabcolsep}{8pt}
\caption{Quasinormal frequencies of the massless scalar field for $\Lambda =0$. For each parameter set the first line is the sixth order WKB value and the second line the Pad\'e averaged value obtained from the same series. All parameters not listed in the first column take the reference values~\eqref{eq:ref} (with $\Lambda =0$).}
\label{tab:wkbs}
\begin{tabular}{llcccc}\hline\hline
 & & \multicolumn{2}{c}{$\ell=1$} & \multicolumn{2}{c}{$\ell=2$}\\ \cline{3-4}\cline{5-6}
 & & $n=0$ & $n=1$ & $n=0$ & $n=1$\\\hline
reference & WKB$_6$ & $0.31607-0.10329i$ & $0.28742-0.32275i$ & $0.52687-0.10181i$ & $0.50678-0.31062i$\\
 & Pad\'e & $0.31609-0.10319i$ & $0.28740-0.32274i$ & $0.52687-0.10180i$ & $0.50678-0.31062i$\\[0.5pt]
$\sigma =0$ & WKB$_6$ & $0.34702-0.11462i$ & $0.31352-0.35937i$ & $0.57464-0.11339i$ & $0.55142-0.34637i$\\
 & Pad\'e & $0.34705-0.11450i$ & $0.31350-0.35935i$ & $0.57464-0.11338i$ & $0.55142-0.34636i$\\[0.5pt]
$\sigma =0.06$ & WKB$_6$ & $0.24967-0.07912i$ & $0.23030-0.24552i$ & $0.42257-0.07746i$ & $0.40877-0.23566i$\\
 & Pad\'e & $0.24968-0.07907i$ & $0.23029-0.24552i$ & $0.42257-0.07745i$ & $0.40877-0.23566i$\\[0.5pt]
$\sigma =0.10$ & WKB$_6$ & $0.17258-0.05175i$ & $0.16204-0.15927i$ & $0.29771-0.05050i$ & $0.29040-0.15294i$\\
 & Pad\'e & $0.17259-0.05173i$ & $0.16203-0.15927i$ & $0.29771-0.05050i$ & $0.29040-0.15294i$\\[0.5pt]
$\alpha =0$ & WKB$_6$ & $0.37597-0.12881i$ & $0.33935-0.40381i$ & $0.62335-0.12690i$ & $0.59723-0.38787i$\\
 & Pad\'e & $0.37599-0.12867i$ & $0.33933-0.40379i$ & $0.62335-0.12689i$ & $0.59723-0.38787i$\\[0.5pt]
$\alpha =0.2$ & WKB$_6$ & $0.25936-0.08021i$ & $0.23782-0.24972i$ & $0.43479-0.07910i$ & $0.41996-0.24088i$\\
 & Pad\'e & $0.25937-0.08015i$ & $0.23782-0.24971i$ & $0.43479-0.07909i$ & $0.41996-0.24088i$\\[0.5pt]
$\alpha =0.3$ & WKB$_6$ & $0.20578-0.05958i$ & $0.19049-0.18473i$ & $0.34707-0.05878i$ & $0.33673-0.17865i$\\
 & Pad\'e & $0.20579-0.05955i$ & $0.19049-0.18473i$ & $0.34707-0.05878i$ & $0.33673-0.17865i$\\[0.5pt]
$q=0$ & WKB$_6$ & $0.30290-0.10144i$ & $0.27341-0.31767i$ & $0.50529-0.09990i$ & $0.48460-0.30507i$\\
 & Pad\'e & $0.30293-0.10131i$ & $0.27340-0.31766i$ & $0.50529-0.09989i$ & $0.48460-0.30507i$\\[0.5pt]
$q=0.6$ & WKB$_6$ & $0.33621-0.10537i$ & $0.30925-0.32805i$ & $0.55985-0.10398i$ & $0.54098-0.31678i$\\
 & Pad\'e & $0.33622-0.10529i$ & $0.30924-0.32804i$ & $0.55985-0.10397i$ & $0.54098-0.31677i$\\[0.5pt]
$q=0.8$ & WKB$_6$ & $0.37646-0.10582i$ & $0.35344-0.32638i$ & $0.62600-0.10473i$ & $0.61010-0.31784i$\\
 & Pad\'e & $0.37646-0.10578i$ & $0.35342-0.32635i$ & $0.62600-0.10472i$ & $0.61010-0.31783i$\\[0.5pt]
$\beta =0.05$ & WKB$_6$ & $0.25357-0.07898i$ & $0.23306-0.24573i$ & $0.42368-0.07790i$ & $0.40942-0.23722i$\\
 & Pad\'e & $0.25358-0.07892i$ & $0.23306-0.24572i$ & $0.42368-0.07789i$ & $0.40942-0.23722i$\\[0.5pt]
$\beta =0.4$ & WKB$_6$ & $0.36935-0.12932i$ & $0.33043-0.40664i$ & $0.61434-0.12722i$ & $0.58661-0.38928i$\\
 & Pad\'e & $0.36939-0.12913i$ & $0.33040-0.40662i$ & $0.61434-0.12721i$ & $0.58661-0.38928i$\\[0.5pt]
$\beta =0.6$ & WKB$_6$ & $0.39662-0.14780i$ & $0.34928-0.46751i$ & $0.65865-0.14508i$ & $0.62428-0.44521i$\\
 & Pad\'e & $0.39668-0.14750i$ & $0.34923-0.46748i$ & $0.65866-0.14505i$ & $0.62427-0.44519i$\\[0.5pt]
$\omega =-0.8$ & WKB$_6$ & $0.30054-0.10002i$ & $0.27610-0.31025i$ & $0.50369-0.09794i$ & $0.48593-0.29820i$\\
 & Pad\'e & $0.30056-0.09994i$ & $0.27609-0.31024i$ & $0.50369-0.09793i$ & $0.48593-0.29820i$\\[0.5pt]
$\omega =-0.5$ & WKB$_6$ & $0.32837-0.10696i$ & $0.29739-0.33500i$ & $0.54526-0.10576i$ & $0.52385-0.32289i$\\
 & Pad\'e & $0.32839-0.10685i$ & $0.29737-0.33499i$ & $0.54526-0.10575i$ & $0.52385-0.32288i$\\[0.5pt]
\hline\hline\end{tabular}\end{table*}

\begin{table*}[tbp]\centering\footnotesize\setlength{\tabcolsep}{8pt}
\caption{Same as Table~\ref{tab:wkbs} for the electromagnetic field.}
\label{tab:wkbem}
\begin{tabular}{llcccc}\hline\hline
 & & \multicolumn{2}{c}{$\ell=1$} & \multicolumn{2}{c}{$\ell=2$}\\ \cline{3-4}\cline{5-6}
 & & $n=0$ & $n=1$ & $n=0$ & $n=1$\\\hline
reference & WKB$_6$ & $0.27398-0.09751i$ & $0.23927-0.30843i$ & $0.50247-0.09980i$ & $0.48102-0.30495i$\\
 & Pad\'e & $0.27404-0.09734i$ & $0.23924-0.30839i$ & $0.50247-0.09979i$ & $0.48102-0.30495i$\\[0.5pt]
$\sigma =0$ & WKB$_6$ & $0.29617-0.10863i$ & $0.25651-0.34465i$ & $0.54504-0.11132i$ & $0.52042-0.34054i$\\
 & Pad\'e & $0.29625-0.10841i$ & $0.25647-0.34460i$ & $0.54504-0.11131i$ & $0.52042-0.34054i$\\[0.5pt]
$\sigma =0.06$ & WKB$_6$ & $0.22419-0.07426i$ & $0.19980-0.23291i$ & $0.40793-0.07578i$ & $0.39308-0.23081i$\\
 & Pad\'e & $0.22423-0.07418i$ & $0.19979-0.23290i$ & $0.40793-0.07577i$ & $0.39308-0.23081i$\\[0.5pt]
$\sigma =0.10$ & WKB$_6$ & $0.16187-0.04871i$ & $0.14875-0.15042i$ & $0.29158-0.04948i$ & $0.28385-0.14985i$\\
 & Pad\'e & $0.16188-0.04868i$ & $0.14874-0.15040i$ & $0.29158-0.04948i$ & $0.28385-0.14985i$\\[0.5pt]
$\alpha =0$ & WKB$_6$ & $0.31986-0.12104i$ & $0.27486-0.38503i$ & $0.59079-0.12418i$ & $0.56273-0.38025i$\\
 & Pad\'e & $0.31998-0.12075i$ & $0.27480-0.38496i$ & $0.59079-0.12417i$ & $0.56273-0.38025i$\\[0.5pt]
$\alpha =0.2$ & WKB$_6$ & $0.22919-0.07609i$ & $0.20349-0.23924i$ & $0.41732-0.07768i$ & $0.40158-0.23683i$\\
 & Pad\'e & $0.22922-0.07600i$ & $0.20348-0.23922i$ & $0.41732-0.07767i$ & $0.40158-0.23682i$\\[0.5pt]
$\alpha =0.3$ & WKB$_6$ & $0.18552-0.05680i$ & $0.16754-0.17747i$ & $0.33535-0.05783i$ & $0.32444-0.17591i$\\
 & Pad\'e & $0.18553-0.05676i$ & $0.16754-0.17746i$ & $0.33535-0.05783i$ & $0.32444-0.17591i$\\[0.5pt]
$q=0$ & WKB$_6$ & $0.26171-0.09542i$ & $0.22593-0.30268i$ & $0.48143-0.09784i$ & $0.45933-0.29924i$\\
 & Pad\'e & $0.26177-0.09528i$ & $0.22591-0.30266i$ & $0.48143-0.09783i$ & $0.45933-0.29924i$\\[0.5pt]
$q=0.6$ & WKB$_6$ & $0.29304-0.09999i$ & $0.26051-0.31482i$ & $0.53479-0.10207i$ & $0.51464-0.31140i$\\
 & Pad\'e & $0.29311-0.09978i$ & $0.26042-0.31468i$ & $0.53479-0.10206i$ & $0.51464-0.31139i$\\[0.5pt]
$q=0.8$ & WKB$_6$ & $0.33249-0.10125i$ & $0.30559-0.31504i$ & $0.60031-0.10302i$ & $0.58350-0.31301i$\\
 & Pad\'e & $0.33256-0.10099i$ & $0.30539-0.31462i$ & $0.60031-0.10301i$ & $0.58349-0.31299i$\\[0.5pt]
$\beta =0.05$ & WKB$_6$ & $0.22260-0.07492i$ & $0.19782-0.23563i$ & $0.40573-0.07649i$ & $0.39051-0.23325i$\\
 & Pad\'e & $0.22263-0.07483i$ & $0.19781-0.23561i$ & $0.40574-0.07649i$ & $0.39051-0.23324i$\\[0.5pt]
$\beta =0.4$ & WKB$_6$ & $0.31461-0.12098i$ & $0.26696-0.38611i$ & $0.58262-0.12433i$ & $0.55292-0.38110i$\\
 & Pad\'e & $0.31475-0.12067i$ & $0.26689-0.38605i$ & $0.58263-0.12431i$ & $0.55292-0.38110i$\\[0.5pt]
$\beta =0.6$ & WKB$_6$ & $0.33216-0.13692i$ & $0.27335-0.44095i$ & $0.62133-0.14130i$ & $0.58439-0.43454i$\\
 & Pad\'e & $0.33239-0.13644i$ & $0.27325-0.44086i$ & $0.62134-0.14128i$ & $0.58438-0.43453i$\\[0.5pt]
$\omega =-0.8$ & WKB$_6$ & $0.26323-0.09356i$ & $0.23149-0.29464i$ & $0.48222-0.09568i$ & $0.46278-0.29184i$\\
 & Pad\'e & $0.26329-0.09341i$ & $0.23146-0.29460i$ & $0.48222-0.09567i$ & $0.46278-0.29184i$\\[0.5pt]
$\omega =-0.5$ & WKB$_6$ & $0.28242-0.10141i$ & $0.24580-0.32122i$ & $0.51853-0.10385i$ & $0.49586-0.31747i$\\
 & Pad\'e & $0.28249-0.10123i$ & $0.24577-0.32118i$ & $0.51853-0.10384i$ & $0.49586-0.31747i$\\[0.5pt]
\hline\hline\end{tabular}\end{table*}

\begin{figure}[tbp]
\centering
\includegraphics[width=\columnwidth]{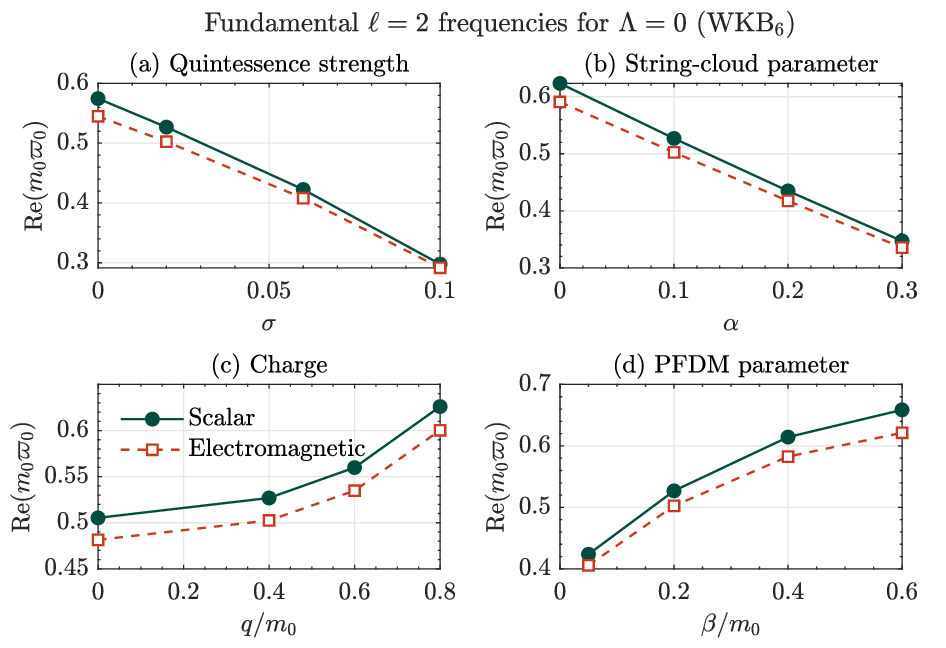}
\caption{Parameter dependence of the fundamental $\ell=2$ oscillation frequency $Re({m}_{0}{\varpi }_{0})$ for $\Lambda =0$, calculated with the sixth-order WKB method. Panels (a)--(d) vary, respectively, the quintessence normalization $\sigma \in \{0,0.02,0.06,0.10\}$, string-cloud parameter $\alpha \in \{0,0.1,0.2,0.3\}$, charge $q/{m}_{0}\in \{0,0.4,0.6,0.8\}$, and PFDM parameter $\beta /{m}_{0}\in \{0.05,0.2,0.4,0.6\}$. All parameters not varied in a panel retain their reference values in Eq.~\eqref{eq:ref}; the four curves are therefore one-parameter scans, not simultaneous variations. Blue circles denote the massless scalar field and vermilion squares the test Maxwell field. The plotted values are the corresponding $\ell=2$, $n=0$ entries of Tables~\ref{tab:wkbs} and \ref{tab:wkbem}.}
\label{fig:scan0}
\end{figure}

The scalar and electromagnetic frequencies are given in Tables~\ref{tab:wkbs} and~\ref{tab:wkbem}, and the fundamental modes are plotted in Fig.~\ref{fig:scan0}.

 All frequencies have ${\rm Im}\,\varpi<0$. Unless stated otherwise, the numbers quoted below are the sixth order values. The electromagnetic modes lie below the scalar ones (for the reference set by about $13\%$ for $\ell=1$ and $5\%$ for $\ell=2$ in the real part), as the term proportional to $f'$ in Eq.~\eqref{eq:V} is absent for $s=1$, but they respond to the parameters in the same way.

The four parameters split into two groups. The quintessence normalization $\sigma$ and the string cloud parameter $\alpha$ decrease both the real part and the magnitude of the imaginary part. The effect is large: when $\sigma$ increases from $0$ to $0.10$, the fundamental scalar $\ell=2$ mode moves from $0.5746-0.1134i$ to $0.2977-0.0505i$, so the real part drops by $48\%$ and the imaginary part by $55\%$. Increasing $\alpha$ from $0$ to $0.3$ has a comparable effect, from $0.6233-0.1269i$ to $0.3471-0.0588i$. The charge $q$ and the PFDM parameter $\beta$ act in the opposite direction and increase both ${\rm Re}\,\varpi$ and $|{\rm Im}\,\varpi|$. Raising $q$ from $0$ to $0.8$ moves the same mode from $0.5053-0.0999i$ to $0.6260-0.1047i$, and raising $\beta$ from $0.05$ to $0.6$ moves it from $0.4237-0.0779i$ to $0.6587-0.1451i$. In the first group the damping rate decreases faster than the oscillation frequency, so the quality factor ${\rm Re}\,\varpi/|{\rm Im}\,\varpi|$ grows. The grouping can be understood from Eq.~\eqref{eq:f}. Near the photon sphere $\alpha$ and $\sigma$ lower $f$ and hence the height of the barrier, while ${q}^{2}/{r}^{2}$ and, for ${r}_{ph}>e|\beta |$, the PFDM term raise it. In terms of the mass function~\eqref{eq:mass}, the string cloud and the quintessence field increase $m(r)$ near the photon sphere, while the charge and the PFDM term decrease it. The charge does so because part of the field energy lies outside $r$, the PFDM term because its energy density is negative for $\beta >0$, Eq.~\eqref{eq:pfdm}. When $\beta$ increases from $0.05$ to $0.6$ the PFDM contribution to $m({r}_{ph})$ changes from $-0.10$ to $-0.40$. The PFDM parameter therefore raises the frequencies, whereas a halo of positive energy density, which increases the enclosed mass, would lower them.

To see the effect of a PFDM term with positive energy density, we also computed the fundamental modes for $\beta =-0.05$, $-0.2$ and $-0.4$, keeping the other parameters at their reference values. The results are listed in Table~\ref{tab:negbeta}.

\begin{table*}[tbp]\centering\footnotesize\setlength{\tabcolsep}{5pt}
\caption{Fundamental modes for $\Lambda =0$ and negative values of the PFDM parameter, for which the PFDM energy density is positive, compared with the reference value $\beta =0.2$. The other parameters take their reference values~\eqref{eq:ref}. $\Delta m=-(\beta /2)\ln ({r}_{ph}/|\beta |)$ is the PFDM contribution to $m({r}_{ph})$. Sixth order WKB values.}
\label{tab:negbeta}
\begin{tabular}{lccccccc}\hline\hline
$\beta$ & ${r}_{+}$ & ${r}_{ph}$ & $\Delta m$ & scalar, $\ell=1$ & scalar, $\ell=2$ & EM, $\ell=1$ & EM, $\ell=2$\\\hline
$0.2$ & 1.7063 & 2.5292 & $-0.254$ & $0.31607-0.10329i$ & $0.52687-0.10181i$ & $0.27398-0.09751i$ & $0.50247-0.09980i$\\
$-0.05$ & 2.5088 & 3.7233 & $+0.108$ & $0.19660-0.05974i$ & $0.32932-0.05890i$ & $0.17408-0.05675i$ & $0.31628-0.05786i$\\
$-0.2$ & 2.9545 & 4.3838 & $+0.309$ & $0.15988-0.04704i$ & $0.26846-0.04636i$ & $0.14288-0.04477i$ & $0.25863-0.04558i$\\
$-0.4$ & 3.3687 & 5.0022 & $+0.505$ & $0.13352-0.03777i$ & $0.22473-0.03723i$ & $0.12047-0.03604i$ & $0.21719-0.03662i$\\
\hline\hline\end{tabular}\end{table*}
 The PFDM contribution to $m({r}_{ph})$ is now positive, the horizon and the photon sphere move outward, and all frequencies decrease. In Ref.~\cite{EslamPanah2026}, $\beta=-0.2$ was likewise found to enlarge the horizon, the photon sphere and the shadow compared with $\beta=0.2$. For the scalar $\ell=2$ mode, $\beta =-0.2$ gives $0.26846-0.04636i$, against $0.52687-0.10181i$ for $\beta =0.2$. For $\beta =-0.2$ we also checked the four modes in the time-domain, with the same settings as in Sec.~\ref{sec:td}; the time-domain and sixth order WKB values agree to within $4\times 10^{-5}$. For these parameters the sign of $\beta$ thus sets the direction of the shift. A PFDM term with positive energy density lowers the frequencies, while the negative energy density obtained for $\beta >0$ raises them.

\subsection{Photon sphere and eikonal limit}

The parameter dependence found numerically can be interpreted geometrically in the large-angular-momentum regime. In that limit, the dominant part of the effective potential is governed by the unstable circular null orbit, so its angular frequency and Lyapunov exponent provide an independent description of the leading oscillation and damping scales.

For large $\ell$ both potentials reduce to $\ell(\ell+1)f/r^2$, whose maximum lies on the circular null orbit, $rf'=2f$. The frequencies then take the form~\cite{Cardoso2009}
\begin{equation}
\begin{aligned}
\varpi\simeq\left(\ell+\tfrac12\right)\Omega_c-i\left(n+\tfrac12\right)\lambda,\qquad
\Omega_c=\frac{\sqrt{f(r_{ph})}}{r_{ph}},\nonumber\\
\lambda=\sqrt{\frac{f(r_{ph})\left[2f(r_{ph})-r_{ph}^2f''(r_{ph})\right]}{2r_{ph}^2}},
\label{eq:eik}
\end{aligned}
\end{equation}
where ${\Omega }_{c}$ and $\lambda$ are the angular velocity and the Lyapunov exponent of the orbit and ${b}_{c}=1/{\Omega }_{c}$ is the critical impact parameter. Note that the photon sphere condition does not contain $\Lambda$, so ${r}_{ph}$ is the same in both regimes; for the reference set ${r}_{ph}=2.5292$~\cite{EslamPanah2026}. For $q=0$ the photon sphere, the shadow and the innermost stable circular orbit of this geometry were analysed in Ref.~\cite{Ahmed2025}. The impact parameter ${b}_{c}$, on the other hand, does depend on $\Lambda$, and the values in Table~\ref{tab:eik} are those for $\Lambda =0$.

\begin{table*}[tbp]\centering\footnotesize\setlength{\tabcolsep}{5pt}
\caption{Horizons and circular null orbit for $\Lambda =0$: event horizon ${r}_{+}$, cosmological horizon ${r}_{c}$ (absent for $\sigma =0$), photon sphere radius ${r}_{ph}$, critical impact parameter ${b}_{c}$, angular velocity ${\Omega }_{c}$ and Lyapunov exponent $\lambda$.}
\label{tab:eik}
\begin{tabular}{lcccccc}\hline\hline
 & ${r}_{+}$ & ${r}_{c}$ & ${r}_{ph}$ & ${b}_{c}$ & ${\Omega }_{c}$ & $\lambda$\\\hline
reference & 1.7063 & 43.956 & 2.5292 & 4.7434 & 0.21082 & 0.20199\\
$\sigma =0$ & 1.6458 & -- & 2.4631 & 4.3660 & 0.22904 & 0.22547\\
$\sigma =0.06$ & 1.8621 & 13.597 & 2.6868 & 5.8638 & 0.17054 & 0.15312\\
$\sigma =0.10$ & 2.1084 & 7.267 & 2.8969 & 8.2423 & 0.12133 & 0.09973\\
$\alpha =0$ & 1.5353 & 49.087 & 2.2815 & 4.0213 & 0.24867 & 0.25169\\
$\alpha =0.2$ & 1.9226 & 38.785 & 2.8400 & 5.7300 & 0.17452 & 0.15699\\
$\alpha =0.3$ & 2.2067 & 33.553 & 3.2439 & 7.1547 & 0.13977 & 0.11670\\
$q=0$ & 1.8057 & 43.952 & 2.6604 & 4.9439 & 0.20227 & 0.19814\\
$q=0.6$ & 1.5638 & 43.961 & 2.3437 & 4.4665 & 0.22389 & 0.20642\\
$q=0.8$ & 1.2954 & 43.969 & 2.0133 & 3.9977 & 0.25014 & 0.20823\\
$\beta =0.05$ & 2.0193 & 43.075 & 3.0015 & 5.8908 & 0.16975 & 0.15460\\
$\beta =0.4$ & 1.5578 & 44.879 & 2.2945 & 4.0729 & 0.24553 & 0.25217\\
$\beta =0.6$ & 1.5331 & 45.660 & 2.2448 & 3.8022 & 0.26301 & 0.28721\\
$\omega =-0.8$ & 1.7225 & 14.189 & 2.5201 & 4.9451 & 0.20222 & 0.19356\\
$\omega =-0.5$ & 1.6916 & 2024.655 & 2.5253 & 4.5941 & 0.21767 & 0.21024\\
\hline\hline\end{tabular}\end{table*}

\begin{figure}[tbp]
\centering
\includegraphics[width=\columnwidth]{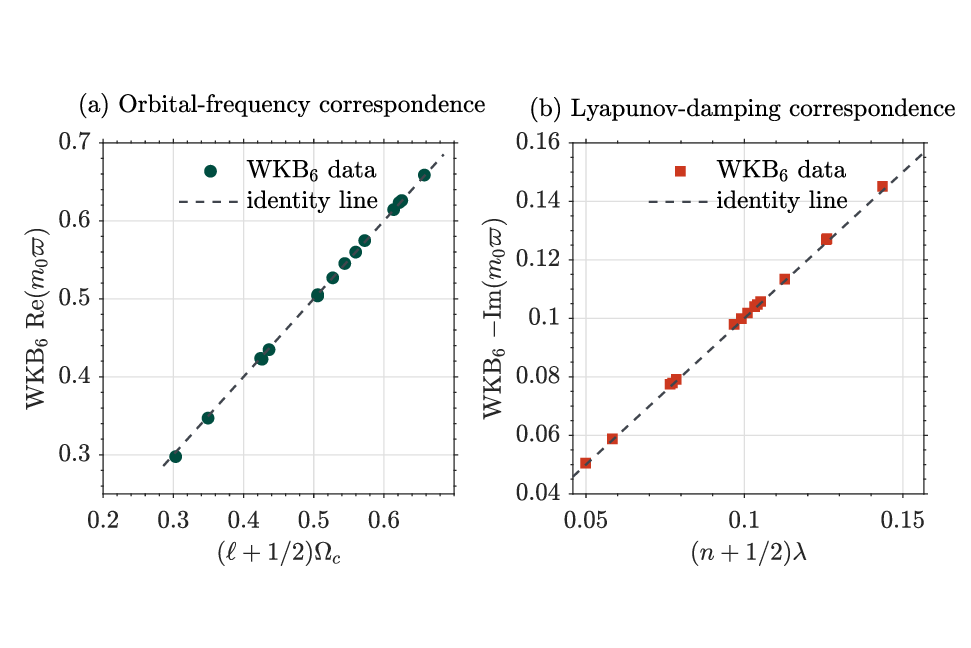}
\caption{Eikonal correspondence between the photon-sphere prediction in Eq.~\eqref{eq:eik} and the sixth-order WKB frequencies for the $\Lambda =0$ parameter sets. Panel (a) compares $Re({m}_{0}\varpi )$ with $(\ell+1/2)m_0\Omega_c$; panel (b) compares $-Im({m}_{0}\varpi )$ with $(n+1/2){m}_{0}\lambda$. The data symbols use the scalar $\ell=2$, $n=0$ entries across the parameter scan in Table~\ref{tab:wkbs}; the dashed diagonal is the identity relation. The photon-sphere quantities $r_{\rm ph}$, ${\Omega }_{c}$, and $\lambda$ are computed from Eq.~\eqref{eq:eik} and the background parameters summarized in Table~\ref{tab:eik}. The residual differences quantify the finite-$\ell$ corrections to the eikonal estimate.}
\label{fig:eikonal}
\end{figure}

The eikonal quantities reproduce the trends of Fig.~\ref{fig:scan0}; the point-by-point comparison with the WKB spectrum is shown in Fig.~\ref{fig:eikonal}.

 When $\sigma$ goes from $0$ to $0.10$, ${\Omega }_{c}$ decreases by $47\%$ and $\lambda$ by $56\%$, practically the same as the changes of ${\rm Re}\,\varpi$ and $|{\rm Im}\,\varpi|$ of the $\ell=2$ mode quoted above. For $\alpha$ from $0$ to $0.3$ the numbers are $44\%$ and $54\%$, again as for the QNMs. For $q$ from $0$ to $0.8$, ${\Omega }_{c}$ grows by $24\%$ and $\lambda$ by only $5\%$, and for $\beta$ from $0.05$ to $0.6$ they grow by $55\%$ and $86\%$, in close agreement with the WKB results. Already for $\ell=2$ the estimate~\eqref{eq:eik} is close to the WKB result. For the reference set it gives $0.52705-0.10100i$, compared with $0.52687-0.10181i$, and for $\sigma =0.10$ it gives $0.30331-0.04987i$, compared with $0.29771-0.05050i$. This suggests that the dependence of the spectrum on the matter fields comes mainly from the unstable null orbit, as expected from the eikonal correspondence~\cite{Cardoso2009}. The eikonal limit also explains why $q$ and $\beta$ behave differently in detail. The charge changes mostly the angular velocity, so its effect on the damping rate is weak, whereas the PFDM parameter changes ${\Omega }_{c}$ and $\lambda$ together.

\section{Time domain analysis for \texorpdfstring{$\Lambda =0$}{Lambda = 0}}
\label{sec:td}

The WKB calculation is based on a local expansion around the potential maximum. To test the resulting frequencies without relying on that approximation, we also evolve the wave equation directly in the time-domain. The late-time signal provides an independent estimate of the fundamental resonance and a check on the parameter trends inferred from the frequency-domain calculation.

As an independent check we integrated the time dependent form of Eq.~\eqref{eq:schr}, $-{\partial }_{t}^{2}\psi +{\partial }_{{r}_{*}}^{2}\psi -V\psi =0$, in light cone coordinates $u=t-{r}_{*}$, $v=t+{r}_{*}$ with the scheme of Gundlach, Price and Pullin~\cite{Gundlach1994},
\begin{equation}
\psi_N=\psi_W+\psi_E-\psi_S-\frac{h^2}{8}\,V_S\left(\psi_W+\psi_E\right)+\mathcal{O}(h^4).
\label{eq:gpp}
\end{equation}
The function $r({r}_{*})$ was obtained by integrating $dr/d{r}_{*}=f$ numerically from $r=3$ (${r}_{*}=0$). We used a Gaussian pulse centred at $v=20$ on the initial surface $u=0$, a grid step $h=0.1$ and the domain $0\leq u,v\leq 1000$, and recorded the signal at ${r}_{*}=10$. The long domain is needed only for the late time analysis below; the ringdown frequencies do not change if it is reduced to $0\leq u,v\leq 500$. The fundamental frequency was extracted from the ringdown with the matrix pencil method~\cite{Hua1990}, using $75\leq t\leq 145$ for $\ell=1$ and $80\leq t\leq 170$ for $\ell=2$. These windows start after the first overtone has decayed, which matters in particular for $q=0.8$. Shifting the windows by $\pm 5$ changes the result by at most $4\times 10^{-4}$.

The comparison is given in Table~\ref{tab:td0} for six parameter sets that cover the full range of Fig.~\ref{fig:scan0}.

\begin{table*}[tbp]\centering\footnotesize\setlength{\tabcolsep}{6pt}
\caption{Fundamental mode for $\Lambda =0$ extracted from the time-domain signals ($\varpi^{\rm TD}$), compared with the sixth order WKB and the Pad\'e averaged values. $\delta_6=|\varpi^{\rm TD}-\varpi^{\rm WKB}_6|/|\varpi^{\rm WKB}_6|$ and $\delta_{\rm P}=|\varpi^{\rm TD}-\varpi^{\rm Pad\acute{e}}|/|\varpi^{\rm Pad\acute{e}}|$, in per cent, computed from the unrounded values.}
\label{tab:td0}
\begin{tabular}{lccccccc}\hline\hline
 & field & $\ell$ & $\varpi^{\rm TD}$ & $\varpi^{\rm WKB}_6$ & $\varpi^{\rm Pad\acute{e}}$ & ${\delta }_{6}$ (\%) & $\delta_{\rm P}$ (\%)\\\hline
reference & scalar & 1 & $0.31609-0.10319i$ & $0.31607-0.10329i$ & $0.31609-0.10319i$ & 0.031 & 0.001\\
 & scalar & 2 & $0.52688-0.10179i$ & $0.52687-0.10181i$ & $0.52687-0.10180i$ & 0.004 & 0.003\\
 & EM & 1 & $0.27405-0.09733i$ & $0.27398-0.09751i$ & $0.27404-0.09734i$ & 0.067 & 0.006\\
 & EM & 2 & $0.50248-0.09979i$ & $0.50247-0.09980i$ & $0.50247-0.09979i$ & 0.003 & 0.002\\[3pt]
$\sigma =0$ & scalar & 1 & $0.34707-0.11449i$ & $0.34702-0.11462i$ & $0.34705-0.11450i$ & 0.039 & 0.006\\
 & scalar & 2 & $0.57467-0.11337i$ & $0.57464-0.11339i$ & $0.57464-0.11338i$ & 0.005 & 0.004\\
 & EM & 1 & $0.29625-0.10841i$ & $0.29617-0.10863i$ & $0.29625-0.10841i$ & 0.075 & 0.001\\
 & EM & 2 & $0.54506-0.11130i$ & $0.54504-0.11132i$ & $0.54504-0.11131i$ & 0.005 & 0.004\\[3pt]
$\sigma =0.10$ & scalar & 1 & $0.17260-0.05177i$ & $0.17258-0.05175i$ & $0.17259-0.05173i$ & 0.017 & 0.026\\
 & scalar & 2 & $0.29772-0.05049i$ & $0.29771-0.05050i$ & $0.29771-0.05050i$ & 0.002 & 0.002\\
 & EM & 1 & $0.16188-0.04869i$ & $0.16187-0.04871i$ & $0.16188-0.04868i$ & 0.015 & 0.002\\
 & EM & 2 & $0.29158-0.04947i$ & $0.29158-0.04948i$ & $0.29158-0.04948i$ & 0.002 & 0.002\\[3pt]
$\alpha =0.3$ & scalar & 1 & $0.20580-0.05949i$ & $0.20578-0.05958i$ & $0.20579-0.05955i$ & 0.046 & 0.030\\
 & scalar & 2 & $0.34707-0.05878i$ & $0.34707-0.05878i$ & $0.34707-0.05878i$ & 0.002 & 0.002\\
 & EM & 1 & $0.18555-0.05671i$ & $0.18552-0.05680i$ & $0.18553-0.05676i$ & 0.051 & 0.028\\
 & EM & 2 & $0.33535-0.05782i$ & $0.33535-0.05783i$ & $0.33535-0.05783i$ & 0.002 & 0.002\\[3pt]
$q=0.8$ & scalar & 1 & $0.37642-0.10576i$ & $0.37646-0.10582i$ & $0.37646-0.10578i$ & 0.017 & 0.010\\
 & scalar & 2 & $0.62602-0.10471i$ & $0.62600-0.10473i$ & $0.62600-0.10472i$ & 0.004 & 0.004\\
 & EM & 1 & $0.33255-0.10099i$ & $0.33249-0.10125i$ & $0.33256-0.10099i$ & 0.077 & 0.002\\
 & EM & 2 & $0.60032-0.10300i$ & $0.60031-0.10302i$ & $0.60031-0.10301i$ & 0.004 & 0.003\\[3pt]
$\beta =0.6$ & scalar & 1 & $0.39650-0.14732i$ & $0.39662-0.14780i$ & $0.39668-0.14750i$ & 0.117 & 0.060\\
 & scalar & 2 & $0.65870-0.14502i$ & $0.65865-0.14508i$ & $0.65866-0.14505i$ & 0.010 & 0.007\\
 & EM & 1 & $0.33243-0.13648i$ & $0.33216-0.13692i$ & $0.33239-0.13644i$ & 0.145 & 0.015\\
 & EM & 2 & $0.62137-0.14126i$ & $0.62133-0.14130i$ & $0.62134-0.14128i$ & 0.009 & 0.006\\
\hline\hline\end{tabular}\end{table*}
 For $\ell=2$ the time-domain frequencies agree with both WKB values to within $0.01\%$. For $\ell=1$ the difference from the sixth order values reaches $0.15\%$, mostly in the imaginary part, while the difference from the Pad\'e values never exceeds $0.06\%$. As in other spacetimes~\cite{Matyjasek2017,Konoplya2019}, the Pad\'e approximants remove most of the residual error of the sixth order formula at low $\ell$.

As an external check we also considered the special case $q=\sigma =\Lambda =0$, $\alpha =0.2$, $\beta =0.3$. This is the Schwarzschild black hole with a cloud of strings in PFDM studied in Ref.~\cite{Hamil2025}, with their parameters $a=0.2$ and $\alpha =0.3$. For the fundamental $\ell=1$ scalar mode, Ref.~\cite{Hamil2025} gives $0.306997-0.101910i$ at sixth WKB order. Our time-domain evolution gives $0.30700-0.10191i$, in agreement to $10^{-5}$. Our sixth order WKB value, $0.30695-0.10206i$, differs by $1.6\times 10^{-4}$, which is within its error estimate ${\Delta }_{6}=2.2\times 10^{-4}$.

\begin{figure}[tbp]
\centering
\includegraphics[width=\columnwidth]{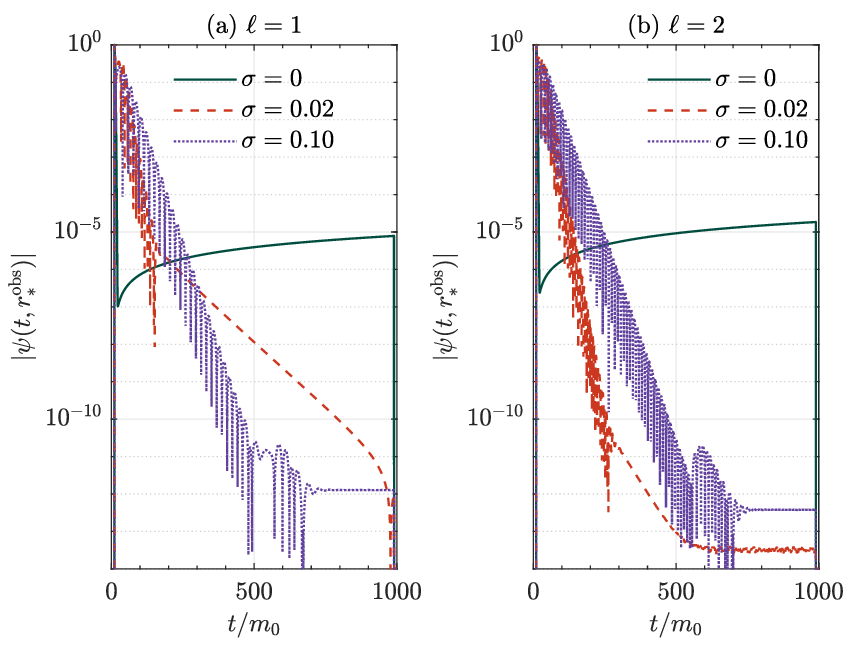}
\caption{Logarithmic time-domain amplitudes $|\psi(t,r_*^{\rm obs})|$ for scalar perturbations in the $\Lambda =0$ geometry, obtained from the characteristic evolution associated with Eq.~\eqref{eq:gpp}. Panels (a) and (b) show $\ell=1$ and $\ell=2$, respectively, at the observer position $r_*^{\rm obs}=10m_0$. The three curves correspond to $\sigma =0$, $0.02$, and $0.10$; all other parameters are fixed to Eq.~\eqref{eq:ref}, with $\Lambda =0$. The characteristic step is $h=0.1$, the initial Gaussian is centered at $v/{m}_{0}=20$, and the computational domain is $0\leq u/{m}_{0},v/{m}_{0}\leq 1000$. The vertical axis is logarithmic and displays the absolute amplitude, not the signed field. Any numerical floor used to avoid plotting zero is a plotting convention and does not represent a physical late-time signal. The comparison is intended to show the change in the ringdown and late-time decay with $\sigma$; quantitative frequencies are reported in Table~\ref{tab:td0}.}
\label{fig:td0}
\end{figure}

Typical logarithmic-amplitude signals are shown in Fig.~\ref{fig:td0}; the decay is slower for larger $\sigma$. For $\sigma =0$ there is no cosmological horizon, and after the ringdown the $\ell=1$ signal enters a slowly decaying tail with an effective exponent of $-4.5$ for $250<t<450$ and $-4.4$ for $500<t<950$. This is close to the Schwarzschild value $-(2\ell+3)=-5$~\cite{Price1972}, but the exponent is still drifting, and since the solid angle deficit and the slowly decaying PFDM term may both change the late time law, we do not quote a tail exponent. For $\sigma =0.02$ a power law does not describe the late signal, while an exponential fit gives decay rates of $0.017$ and $0.011$ in the same two windows. This resembles the Schwarzschild--de Sitter case, where the tail is replaced by an exponential decay set by the surface gravity of the cosmological horizon~\cite{Brady1997}. The measured rate is of the same order as ${\kappa }_{c}=0.0097$ (Sec.~\ref{sec:geometry}), but our evolutions are too short to fix the asymptotic rate. For $\ell=2$ the tail falls below the roundoff level ($|\psi |\sim 10^{-11}$).

\section{Quasinormal modes in AdS: spectral method}
\label{sec:ads}

For asymptotically AdS spacetimes, the conformal boundary is timelike and the scalar evolution requires a boundary condition there; it is not described by the purely outgoing condition used at a cosmological horizon~\cite{Horowitz2000,IshibashiWald2004,ChanMann1997}. The quasinormal frequencies are consequently determined by a global boundary-value problem, with an ingoing condition at the future event horizon and the normalizable branch selected at the AdS boundary. We implement the horizon condition by using ingoing Eddington--Finkelstein coordinates and discretize the compactified radial equation with Chebyshev--Gauss--Lobatto collocation~\cite{Trefethen2000,Jansen2017}.

\subsection{Setting up the eigenvalue problem}

The numerical construction is organized so that the horizon regularity and the AdS boundary condition are built into the radial eigenvalue problem. We first establish the exterior domain, then derive the regular radial equation and specify the endpoint conditions before discretizing the operator.

For $\Lambda <0$ we adopt the ingoing-coordinate spectral strategy used in numerical studies of black-hole quasinormal modes~\cite{Jansen2017}; the compactification and collocation follow standard Chebyshev spectral methods~\cite{Trefethen2000}. The AdS boundary conditions themselves are those of the scalar-field problem in asymptotically AdS spacetimes~\cite{Horowitz2000,IshibashiWald2004}. We give the construction explicitly because the horizon condition, the conformal-boundary condition, and the discretized differential operator together define the eigenvalue problem.

\emph{\textbf{Exterior region}.} For each parameter set we first fix $f(r)$, $\Lambda$ and $\ell$, and find the event horizon ${r}_{+}$ as the largest positive root of $f({r}_{+})=0$. For the reference set the lapse function has two positive roots, an inner horizon at ${r}_{-}=0.0752$ and the event horizon at ${r}_{+}=1.6556$; for $q=0$ only one positive root remains. We then checked numerically that $f(r)>0$ for all $r>{r}_{+}$, so that the exterior is static all the way to the AdS boundary, where $f\simeq |\Lambda |{r}^{2}/3$. This holds throughout the AdS parameter range examined.

\emph{\textbf{Radial equation}.} The massless scalar field obeys
\begin{equation}
\Box\Phi=\frac{1}{\sqrt{-g}}\,\partial_\mu\!\left(\sqrt{-g}\,g^{\mu\nu}\partial_\nu\Phi\right)=0 .
\end{equation}
We use the ingoing Eddington--Finkelstein coordinate $v=t+{r}_{*}$, in which the line element~\eqref{eq:metric} becomes
\begin{equation}
ds^2=-f(r)\,dv^2+2\,dv\,dr+r^2\left(d\theta^2+\sin^2\theta\,d\varphi^2\right)
\end{equation}
and is regular at the horizon. The nonzero inverse components are ${g}^{vr}=1$, ${g}^{rr}=f$, ${g}^{\theta \theta }={r}^{-2}$ and ${g}^{\varphi \varphi }=(r\sin \theta {)}^{-2}$, and $\sqrt{-g}={r}^{2}\sin \theta$. We separate the time and angular dependence as
\begin{equation}
\Phi=e^{-i\varpi v}\,Y_{\ell m}(\theta,\varphi)\,R(r),
\end{equation}
where the complex frequency $\varpi$ is the unknown of the problem. The wave equation then reduces to
\begin{equation}
\frac{d}{dr}\!\left(r^2f\,\frac{dR}{dr}\right)-2i\varpi\left(r^2\frac{dR}{dr}+rR\right)-\ell(\ell+1)R=0 .
\label{eq:radEF}
\end{equation}
This can be compared with the separation of Sec.~\ref{sec:pert}, $R={e}^{i\varpi {r}_{*}}\psi /r$.

\emph{\textbf{Compactification}.} Next we introduce $z=1/r$, which maps the exterior ${r}_{+}\leq r<\infty$ onto $0<z\leq {z}_{+}=1/{r}_{+}$ and places the AdS boundary at $z=0$, and define
\begin{equation}
H(z)=z^2f(1/z).
\end{equation}
For the lapse function~\eqref{eq:f},
\begin{equation}
\begin{aligned}
H(z)={}&-\frac{\Lambda}{3}+(1-\alpha)z^2-2m_0z^3+q^2z^4\\
&+\beta z^3\ln\frac{1}{|\beta|z}-\sigma z^{3(1+\omega)}.
\end{aligned}
\label{eq:H}
\end{equation}
so that $H(0)=|\Lambda |/3>0$ and $H({z}_{+})=0$. Writing $R={z}^{3}\phi (z)$, Eq.~\eqref{eq:radEF} becomes
\begin{equation}
\begin{aligned}
zH\,\phi''+\left(4H+zH'\right)\phi'
+\left[3H'-\ell(\ell+1)z\right]\phi\\
=-2i\varpi\left(z\phi'+2\phi\right).
\end{aligned}
\label{eq:spec}
\end{equation}
where a prime now denotes $d/dz$. In these variables the frequency enters only linearly, which is the reason for using Eddington--Finkelstein coordinates.

\emph{\textbf{Boundary conditions}.} Both endpoints are regular singular points of Eq.~\eqref{eq:spec}. Near the AdS boundary the leading terms are $H(0)\left({z\phi ''+4\phi '}\right)$, and the solutions behave as $\phi \sim {z}^{s}$ with $s(s+3)=0$. For the field itself this means $\Phi \sim a+b/{r}^{3}$ at large $r$. The branch $s=-3$ carries the nonnormalizable coefficient $a$ and the branch $s=0$ the normalizable coefficient $b$. As in the standard AdS scalar problem~\cite{Horowitz2000,Batic2026,ChanMann1997}, we impose the Dirichlet condition $a=0$, which for a massless scalar is the boundary condition selected by requiring a finite energy~\cite{IshibashiWald2004}. Since the factor ${z}^{3}$ has been extracted, this condition simply says that $\phi$ is regular at $z=0$. Other choices, such as a linear relation between $a$ and $b$, define a different spectral problem; for the massless scalar they are not associated with a finite energy, and they can produce formal eigenvalues with ${\rm Im}\,\varpi>0$~\cite{Batic2026}. We do not consider them here. At the horizon $H$ vanishes linearly, $H\simeq 2\kappa ({z}_{+}-z)$ with the surface gravity $\kappa =f'({r}_{+})/2$ ($\kappa =0.2955$ for the reference set), and the exponents are $s=0$ and $s=i\varpi /\kappa$. The second branch, $\phi \sim ({z}_{+}-z{)}^{i\varpi /\kappa }$, corresponds to $\Phi \propto {e}^{-i\varpi (t-{r}_{*})}$, a wave coming out of the horizon, and is excluded. The regular branch is the ingoing one. Both QNM conditions are therefore equivalent to the requirement that $\phi$ be smooth on the closed interval $0\leq z\leq {z}_{+}$, and no boundary condition has to be added by hand. At the endpoints Eq.~\eqref{eq:spec} itself reduces to the relations that a regular solution must satisfy; at $z=0$, for example, it gives $4H(0)\phi '(0)+3H'(0)\phi (0)=-4i\varpi \phi (0)$.

\emph{\textbf{Discretization}.} We map $0\leq z\leq {z}_{+}$ onto $-1\leq x\leq 1$ by $z={z}_{+}(1-x)/2$ and represent $\phi$ by its values at the $N+1$ Chebyshev--Gauss--Lobatto points ${x}_{j}=\cos (j\pi /N)$, $j=0,\dots,N$. Derivatives are replaced by the Chebyshev differentiation matrix $D$~\cite{Trefethen2000}, and Eq.~\eqref{eq:spec} is imposed at all $N+1$ points, including the two endpoints. This gives the generalized eigenvalue problem
\begin{equation}
\mathbf{A}\bm{\phi}=\varpi\,\mathbf{B}\bm{\phi},
\end{equation}
with
\begin{equation}
\begin{aligned}
\mathbf{A}&={\rm diag}(zH)\,\mathbf{D}^2+{\rm diag}(4H+zH')\,\mathbf{D}\nonumber\\
&+{\rm diag}\big(3H'-\ell(\ell+1)z\big),\\
\mathbf{B}&=-2i\left[{\rm diag}(z)\,\mathbf{D}+2\,\mathbf{I}\right],
\end{aligned}
\end{equation}
where the functions are evaluated on the grid. The problem is solved directly for all $N+1$ eigenvalues.

\emph{\textbf{Selection of the modes and convergence}.} Only part of the eigenvalues approximate QNMs; the others change with $N$ and are spurious. We therefore solved the problem with $N=50$ and $N=70$ and kept only those eigenvalues that agree to a relative accuracy of $10^{-6}$. Because some imaginary parts are very small, we also repeated the calculation with $N=90$ and $N=110$. All computed imaginary parts, including the smallest one, $-1.31\times 10^{-7}$, are stable to at least three significant digits. Table~\ref{tab:conv} shows how the fundamental modes of the reference set converge.

\begin{table}[tbp]\centering\scriptsize\setlength{\tabcolsep}{6pt}
\caption{Convergence of the fundamental scalar modes of the reference AdS black hole with the number of grid points. Values with ${\rm Im}\,\varpi>0$ are numerical artefacts of the coarse grids.}
\label{tab:conv}
\begin{tabular}{ccc}\hline\hline
$N$ & $\ell=1$ & $\ell=2$\\\hline
10 & $0.35168383+1.2091\times 10^{-2}i$ & $0.38414400-3.4848\times 10^{-2}i$\\
15 & $0.32755729-6.4260\times 10^{-3}i$ & $0.42768244+5.0305\times 10^{-3}i$\\
20 & $0.32555288-3.0985\times 10^{-3}i$ & $0.43167203+2.9481\times 10^{-4}i$\\
30 & $0.32538236-2.9277\times 10^{-3}i$ & $0.43165143-8.4450\times 10^{-5}i$\\
40 & $0.32538154-2.9272\times 10^{-3}i$ & $0.43165208-8.5765\times 10^{-5}i$\\
50 & $0.32538153-2.9272\times 10^{-3}i$ & $0.43165208-8.5770\times 10^{-5}i$\\
70 & $0.32538153-2.9272\times 10^{-3}i$ & $0.43165208-8.5770\times 10^{-5}i$\\
\hline\hline\end{tabular}\end{table}
 For $N=10$ the $\ell=1$ eigenvalue still has a positive imaginary part, and so does the $\ell=2$ eigenvalue for $N=15$ and $N=20$. This is a numerical artefact, and it disappears when the grid is refined. From $N=40$ on, the $\ell=1$ frequency changes by less than $10^{-7}$. The $\ell=2$ mode, whose damping rate is below $10^{-4}$, needs a few more points, and from $N=50$ on its imaginary part is stable to five significant digits.

\subsection{Tests}

Before applying the spectral scheme to the matter-supported geometry, we test it on Schwarzschild--AdS, for which scalar quasinormal frequencies have been computed using the Horowitz--Hubeny power-series method and subsequent high-accuracy numerical approaches~\cite{BertiKokkotas2003,Horowitz2000,Batic2026,Konoplya2002}. This benchmark isolates the numerical implementation from the additional environmental parameters and checks the multipoles used below.

We tested the code on the Schwarzschild--AdS black hole, $f=1-2M/r+{r}^{2}/{L}^{2}$ with $L=1$, for three horizon radii and $\ell=0,1,2$. Table~\ref{tab:sads} compares our fundamental modes with the power series results of Horowitz and Hubeny~\cite{Horowitz2000}, available for $\ell=0$, and with the high precision spectral results of Batic, Cornell and Dutykh~\cite{Batic2026}.

\begin{table}[tbp]\centering\scriptsize\setlength{\tabcolsep}{3pt}
\caption{Fundamental scalar modes of the Schwarzschild--AdS black hole with $L=1$, used to test the spectral code. HH: Horowitz and Hubeny~\cite{Horowitz2000}; BCD: Batic, Cornell and Dutykh~\cite{Batic2026}.}
\label{tab:sads}
\begin{tabular}{ccccc}\hline\hline
${r}_{+}$ & $\ell$ & this work & HH & BCD\\\hline
0.4 & 0 & $2.3629-1.0065i$ & $2.3629-1.0064i$ & $2.3629-1.0065i$\\
0.4 & 1 & $3.1338-0.7029i$ & -- & $3.1338-0.7029i$\\
0.4 & 2 & $4.1285-0.4512i$ & -- & $4.1285-0.4512i$\\
1 & 0 & $2.7982-2.6712i$ & $2.7982-2.6712i$ & $2.7982-2.6712i$\\
1 & 1 & $3.3306-2.4895i$ & -- & $3.3306-2.4895i$\\
1 & 2 & $4.1171-2.2756i$ & -- & $4.1171-2.2756i$\\
10 & 0 & $18.6070-26.6418i$ & $18.6070-26.6418i$ & $18.6070-26.6418i$\\
\hline\hline\end{tabular}\end{table}
 All entries agree to the four decimals shown, apart from the last digit of one $\ell=0$ value of Ref.~\cite{Horowitz2000}, where our result coincides with that of Ref.~\cite{Batic2026}. The tests with $\ell=1$ and $\ell=2$ matter here, because these are the multipoles used below. For ${r}_{+}=1$ and $\ell=0$ we also find the first overtone, $4.7585-5.0376i$, again as in Ref.~\cite{Batic2026}.

For the metric~\eqref{eq:f} the PFDM term contributes $\beta {z}^{3}\ln z$ to $H$, which is not analytic at $z=0$. It vanishes there together with its first derivative, and the eigenvalues still converge, as the agreement between the different grids shows. The situation is different for $\omega =-0.8$ and $\omega =-0.5$, where the quintessence term in Eq.~\eqref{eq:H} is $-\sigma {z}^{0.6}$ and $-\sigma {z}^{1.5}$. In the first case $H'\propto {z}^{-0.4}$ diverges at the boundary, and in the second $H'$ contains the nonanalytic term ${z}^{1/2}$. In both cases the Chebyshev expansion converges poorly. For these two parameter sets we therefore took the frequencies from the time-domain evolutions described below.

\subsection{Time domain cross check}

A second validation is obtained by evolving the AdS wave equation directly. Unlike the spectral calculation, this approach follows the field in time and therefore tests whether the small imaginary parts of the computed eigenfrequencies are reflected in the decay of the signal.

The time-domain code integrates ${\partial }_{t}^{2}\psi ={\partial }_{{r}_{*}}^{2}\psi -V\psi$ on a uniform grid in ${r}_{*}$ with a second order leapfrog scheme ($\Delta {r}_{*}=0.05$, $\Delta t=0.025$). A Dirichlet condition is imposed at the AdS boundary ${r}_{*}={r}_{*}^{\infty }$, and a simple outgoing condition at ${r}_{*}=-2200$, far enough that nothing reflected there reaches the observer before the end of the run at $t=2000$. The initial data are a static Gaussian of width $1.5$ centred at ${r}_{*}=-3$, and the signal is recorded at ${r}_{*}=5$. For the evaluation of $V({r}_{*})$ at large $r$ the tortoise coordinate was computed by integrating in $z$ from $z=1/3$ to $z=0$, which gives ${r}_{*}^{\infty }$ accurately. The two longest-lived modes were extracted with the matrix pencil method from the interval $200\leq t\leq 1400$, and the interval $400\leq t\leq 1600$ was used as a check.

The comparison for the reference set is shown in Table~\ref{tab:adstd}.

\begin{figure}[tbp]
\centering
\includegraphics[width=\columnwidth]{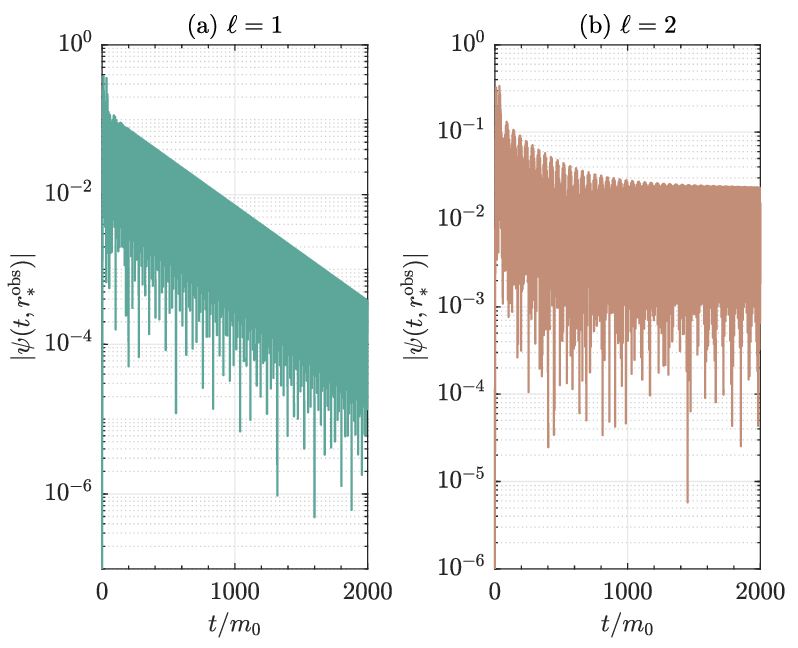}
\caption{Logarithmic time-domain amplitudes $|\psi(t,r_*^{\rm obs})|$ for massless scalar perturbations of the reference AdS black hole, with $\Lambda =-0.03$ and the remaining parameters fixed by Eq.~\eqref{eq:ref}. Panels (a) and (b) correspond to $\ell=1$ and $\ell=2$, respectively. The evolution solves ${\partial }_{t}^{2}\psi ={\partial }_{{r}_{*}}^{2}\psi -V\psi$ with $\Delta {r}_{*}/{m}_{0}=0.05$, $\Delta t/{m}_{0}=0.025$, and ${t}_{\max\limits_{}}/{m}_{0}=2000$, using a Gaussian initial profile of width $1.5{m}_{0}$ centered at ${r}_{*}=-3{m}_{0}$, zero initial velocity, a Dirichlet condition at the AdS boundary, and an outgoing condition at the inner numerical boundary ${r}_{*}=-2200{m}_{0}$. The signal is recorded at $r_*^{\rm obs}=5m_0$. The vertical axis is logarithmic and represents the absolute field amplitude. The very slow decay for $\ell=2$ is consistent with the small imaginary part of the fundamental mode obtained from the spectral calculation.}
\label{fig:tdads}
\end{figure}

\begin{table}[tbp]\centering\scriptsize\setlength{\tabcolsep}{6pt}
\caption{The two longest-lived scalar modes of the reference AdS black hole: spectral method versus time-domain evolution.}
\label{tab:adstd}
\begin{tabular}{cccc}\hline\hline
$\ell$ & $n$ & spectral & time-domain\\\hline
0 & 0 & $0.22072-0.02444i$ & $0.22072-0.02444i$ \\
0 & 1 & $0.36499-0.06790i$ & $0.36500-0.06792i$ \\
1 & 0 & $0.32538-0.00293i$ & $0.32538-0.00293i$ \\
1 & 1 & $0.45056-0.02754i$ & $0.45056-0.02754i$ \\
2 & 0 & $0.43165-8.58\times 10^{-5}i$ & $0.43165-8.58\times 10^{-5}i$ \\
2 & 1 & $0.56535-0.00382i$ & $0.56535-0.00382i$ \\
\hline\hline\end{tabular}\end{table}
 The same comparison was repeated with a coarser grid ($\Delta {r}_{*}=0.1$, $\Delta t=0.05$, $t\leq 1000$), and a shorter fitting window ($200\leq t\leq 900$). This changes the real parts by less than $3\times 10^{-5}$ and the imaginary parts by less than $0.2\%$ (relative); the same holds for the two sets with $\omega =-0.8$ and $\omega =-0.5$. The two methods agree to four or more significant digits in the real parts and to within $3\times 10^{-4}$ (relative) in the imaginary parts, even for the $\ell=2$ fundamental mode, whose imaginary part is only $-8.6\times 10^{-5}$. Figure~\ref{fig:tdads} shows the corresponding logarithmic-amplitude signals.

 The late-time signal is a superposition of a few very weakly damped oscillations; on the logarithmic-amplitude scale in Fig.~\ref{fig:tdads}, the $\ell=2$ envelope changes only weakly over the available run. This is the exponential late time decay characteristic of AdS black holes~\cite{Horowitz2000,Wang2001}.

We repeated the time-domain comparison for all AdS parameter sets considered here and $\ell=0,1,2$, with the same numerical settings. The modes were extracted with the matrix pencil method in two pairs of windows, $200\leq t\leq 1400$ and $400\leq t\leq 1600$ for the weakly damped modes and $80\leq t\leq 600$ and $120\leq t\leq 640$ for the more strongly damped ones, and a mode was accepted only if both windows of a pair gave the same frequency. For 44 of the 45 fundamental modes the time-domain and spectral frequencies agree to better than $3\times 10^{-6}$ in the real part and $6\times 10^{-4}$ (relative) in the imaginary part, including damping rates as small as $6\times 10^{-6}$ ($q=0.8$, $\ell=2$). All 45 second modes are recovered as well, with differences below $10^{-4}$ in the real part and, apart from one case, below $0.3\%$ in the imaginary part. The exceptions are the two $\ell=2$ modes for $\Lambda =-0.01$, with imaginary parts $-1.31\times 10^{-7}$ and $-4.37\times 10^{-6}$. Their decay times, of order $10^{7}$ and $2\times 10^{5}$, are much longer than our evolutions, so they cannot be resolved in the time-domain, while the spectral method gives them without difficulty.

\subsection{Spectrum}

With the boundary-value formulation and its independent checks in place, we turn to the dependence of the AdS spectrum on the black-hole and environmental parameters. The results below emphasize the two longest-lived modes for each angular sector, since these modes most clearly reveal the effect of the reflecting boundary and the potential barrier.

\begin{table*}[tbp]\centering\small\setlength{\tabcolsep}{10pt}\renewcommand{\arraystretch}{1.05}
\caption{The two longest-lived scalar quasinormal modes of the AdS black hole for $\ell=0,1,2$. For each parameter set the first line is the fundamental mode and the second line the next mode. Parameters not listed take the reference values~\eqref{eq:ref}. Rows marked with $^\dagger$ were obtained from time-domain evolutions (see text); all other rows are spectral results.}
\label{tab:ads}
\begin{tabular}{lcccc}\hline\hline
 & ${r}_{+}$ & $\ell=0$ & $\ell=1$ & $\ell=2$\\\hline
reference & 1.6556 & $0.22072-0.02444i$ & $0.32538-0.00293i$ & $0.43165-8.58\times 10^{-5}i$\\
 &  & $0.36499-0.06790i$ & $0.45056-0.02754i$ & $0.56535-0.00382i$\\[0.5pt]
$\sigma =0$ & 1.6030 & $0.24609-0.02645i$ & $0.35403-0.00339i$ & $0.46485-1.07\times 10^{-4}i$\\
 &  & $0.40234-0.07361i$ & $0.49072-0.03074i$ & $0.61084-0.00452i$\\[0.5pt]
$\sigma =0.06$ & 1.7855 & $0.16913-0.02007i$ & $0.26573-0.00200i$ & $0.36135-4.91\times 10^{-5}i$\\
 &  & $0.28768-0.05546i$ & $0.36656-0.02076i$ & $0.46903-0.00246i$\\[0.5pt]
$\sigma =0.10$ & 1.9699 & $0.11586-0.01499i$ & $0.20108-0.00109i$ & $0.28268-2.06\times 10^{-5}i$\\
 &  & $0.20528-0.04101i$ & $0.27518-0.01329i$ & $0.36176-0.00119i$\\[0.5pt]
$\alpha =0$ & 1.5015 & $0.23834-0.01955i$ & $0.34335-0.00166i$ & $0.44803-3.29\times 10^{-5}i$\\
 &  & $0.39173-0.05855i$ & $0.48195-0.01889i$ & $0.59853-0.00147i$\\[0.5pt]
$\alpha =0.2$ & 1.8429 & $0.20259-0.03053i$ & $0.30560-0.00540i$ & $0.41345-2.80\times 10^{-4}i$\\
 &  & $0.33753-0.07856i$ & $0.41854-0.03892i$ & $0.52830-0.00962i$\\[0.5pt]
$\alpha =0.3$ & 2.0741 & $0.18409-0.03801i$ & $0.28401-0.00996i$ & $0.39250-0.00110i$\\
 &  & $0.30930-0.09083i$ & $0.38636-0.05309i$ & $0.48874-0.02122i$\\[0.5pt]
$q=0$ & 1.7489 & $0.22130-0.02601i$ & $0.32438-0.00375i$ & $0.43103-1.43\times 10^{-4}i$\\
 &  & $0.36705-0.07052i$ & $0.44994-0.03134i$ & $0.56274-0.00552i$\\[0.5pt]
$q=0.6$ & 1.5207 & $0.21983-0.02234i$ & $0.32674-0.00196i$ & $0.43240-3.80\times 10^{-5}i$\\
 &  & $0.36187-0.06455i$ & $0.45166-0.02234i$ & $0.56885-0.00206i$\\[0.5pt]
$q=0.8$ & 1.2626 & $0.21785-0.01912i$ & $0.32886-7.85\times 10^{-4}i$ & $0.43341-5.93\times 10^{-6}i$\\
 &  & $0.35576-0.06052i$ & $0.45418-0.01389i$ & $0.57395-4.68\times 10^{-4}i$\\[0.5pt]
$\beta =0.05$ & 1.9300 & $0.20963-0.03393i$ & $0.31129-0.00718i$ & $0.42039-5.00\times 10^{-4}i$\\
 &  & $0.34845-0.08636i$ & $0.42762-0.04621i$ & $0.53582-0.01404i$\\[0.5pt]
$\beta =0.4$ & 1.5235 & $0.23254-0.01894i$ & $0.33680-0.00161i$ & $0.44072-3.30\times 10^{-5}i$\\
 &  & $0.38417-0.05596i$ & $0.47343-0.01797i$ & $0.58832-0.00140i$\\[0.5pt]
$\beta =0.6$ & 1.5039 & $0.24159-0.01688i$ & $0.34427-0.00138i$ & $0.44709-2.83\times 10^{-5}i$\\
 &  & $0.39983-0.05098i$ & $0.48918-0.01524i$ & $0.60258-0.00104i$\\[0.5pt]
$\omega =-0.8$$^{\dagger}$ & 1.6686 & $0.18673-0.01611i$ & $0.28057-0.00113i$ & $0.37231-1.84\times 10^{-5}i$\\
 &  & $0.31252-0.04855i$ & $0.39350-0.01434i$ & $0.49526-9.32\times 10^{-4}i$\\[0.5pt]
$\omega =-0.5$$^{\dagger}$ & 1.6435 & $0.23669-0.02735i$ & $0.34413-0.00372i$ & $0.45472-1.27\times 10^{-4}i$\\
 &  & $0.38845-0.07460i$ & $0.47555-0.03232i$ & $0.59405-0.00528i$\\[0.5pt]
$\Lambda =-0.01$ & 1.6884 & $0.12575-0.00407i$ & $0.18569-4.45\times 10^{-5}i$ & $0.24170-1.31\times 10^{-7}i$\\
 &  & $0.20744-0.01478i$ & $0.26826-7.50\times 10^{-4}i$ & $0.32769-4.37\times 10^{-6}i$\\[0.5pt]
$\Lambda =-0.05$ & 1.6263 & $0.28370-0.04912i$ & $0.41112-0.01322i$ & $0.55322-0.00143i$\\
 &  & $0.47259-0.12213i$ & $0.57020-0.07227i$ & $0.70516-0.02785i$\\[0.5pt]
\hline\hline
\end{tabular}
\end{table*}

\begin{figure*}[tbp]
\centering
\includegraphics[width=\textwidth]{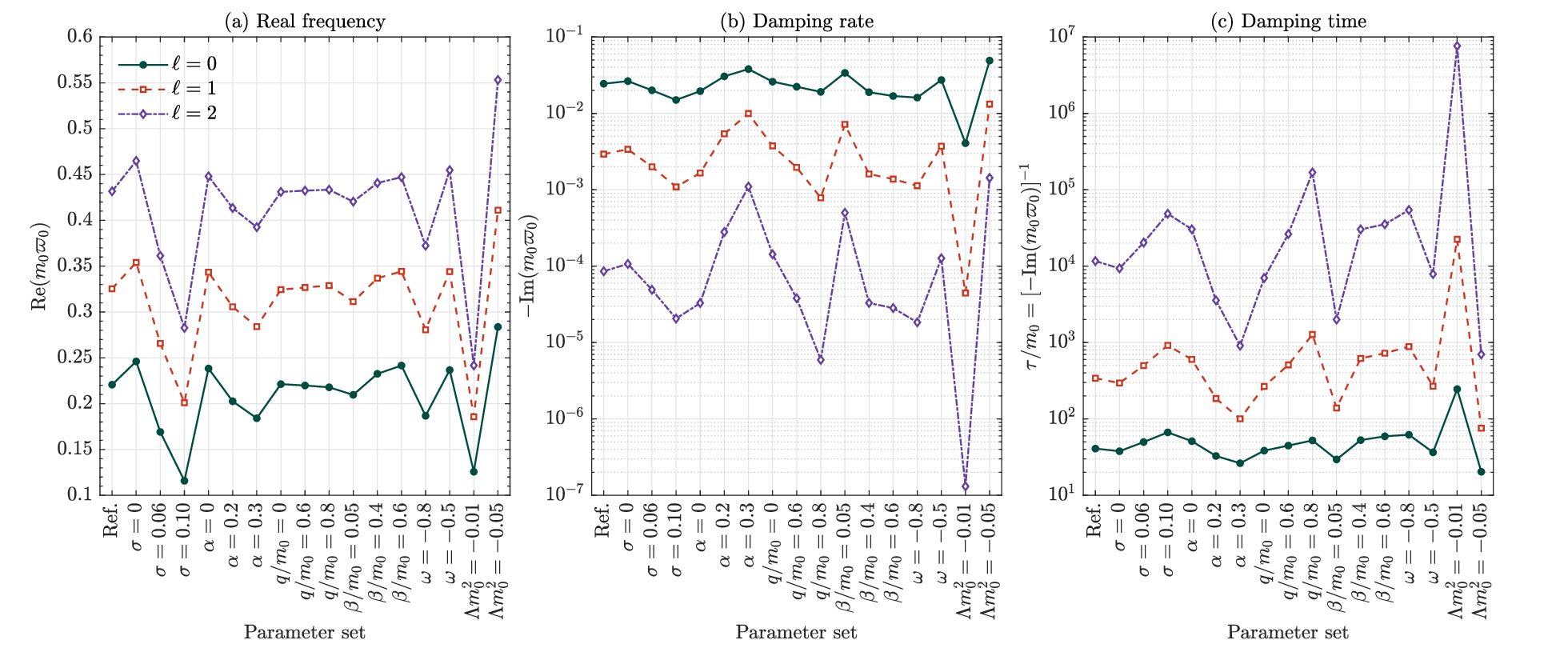}
\caption{Fundamental scalar spectrum in the AdS backgrounds listed in Table~\ref{tab:ads}. The horizontal coordinate labels the parameter set, with all non-listed parameters fixed to Eq.~\eqref{eq:ref}. Panel (a) shows $Re({m}_{0}{\varpi }_{0})$ for $\ell=0,1,2$; panel (b) shows the corresponding damping rates $-Im({m}_{0}{\varpi }_{0})$ on a logarithmic scale; and panel (c) displays the dimensionless damping times $\tau /{m}_{0}=[-Im({m}_{0}{\varpi }_{0}){]}^{-1}$, also logarithmically. The $\omega =-0.8$ and $\omega =-0.5$ entries are the time-domain results described in Sec.~\ref{sec:ads}; all other plotted values are from the Chebyshev spectral calculation. Only the fundamental ($n=0$) entries are plotted; the next modes listed in the table are not included.}
\label{fig:adsspectrum}
\end{figure*}

The two longest-lived scalar modes for $\ell=0,1,2$ are collected in Table~\ref{tab:ads}. Their fundamental branches across the parameter scan are compared in Fig.~\ref{fig:adsspectrum}.

The logarithmic damping and lifetime panels expose the multi-decade variation without obscuring the real-frequency trends. For the reference parameters the fundamental $\ell=1$ mode is $0.32538-0.00293i$ in AdS and $0.31607-0.10329i$ for $\Lambda =0$. The real parts are similar, but the damping rate in AdS is smaller by a factor of about 35, and for $\ell=2$ the factor exceeds $10^{3}$. The modes are long lived because the wave is reflected by the AdS wall and can escape into the horizon only by tunnelling through the barrier. We come back to this point in Sec.~\ref{sec:cavity}. All modes found have ${\rm Im}\,\varpi<0$, so there is no sign of an instability in the parameter range studied.

The dependence of the real parts on the matter parameters is similar to that for $\Lambda =0$. Larger $\sigma$ and $\alpha$ lower ${\rm Re}\,\varpi$, while larger $q$ and $\beta$ raise it; for $\ell=1$ the increase is about $1\%$ for $q$ from $0$ to $0.8$ and about $10\%$ for $\beta$ from $0.05$ to $0.6$. The cosmological constant has the strongest effect. For $\ell=1$, ${\rm Re}\,\varpi_0$ takes the values $0.1857$, $0.3254$ and $0.4111$ for $\Lambda =-0.01,-0.03,-0.05$, approximately in the ratio $1:\sqrt{3}:\sqrt{5}$, i.e.\ ${\rm Re}\,\varpi_0\propto\sqrt{|\Lambda|}$. The same scaling holds for pure AdS space, where the normal modes of a massless scalar are $\varpi L=\ell+2n+3$ with $L=\sqrt{3/|\Lambda |}$. For $\ell=1$ our fundamental frequencies are about $0.80$ times the pure AdS value $4/L$ for all three values of $\Lambda$, which suggests that the trapped modes are close to AdS normal modes, shifted down by the gravitational redshift of the black hole and its environment, as also found for small Schwarzschild--AdS black holes, whose scalar modes approach the pure AdS values when ${r}_{+}/L\rightarrow 0$~\cite{Batic2026}.

The imaginary parts behave quite differently. Here the parameters do not simply shift the frequency but change the damping rate by orders of magnitude. Increasing $\alpha$ from $0$ to $0.3$ increases $|{\rm Im}\,\varpi_0|$ for $\ell=1$ by a factor of six, increasing $q$ from $0$ to $0.8$ decreases it by a factor of five, and changing $\Lambda$ from $-0.01$ to $-0.05$ increases it by more than two orders of magnitude. We discuss this behaviour in the next section.

\section{Barrier, cavity and the damping of trapped modes}
\label{sec:cavity}

The long-lived modes in the AdS spectrum can be interpreted in terms of partial wave trapping between the potential barrier associated with the photon sphere and the reflecting conformal boundary. In the geometric-optics and large-angular-momentum regimes, the interplay between the photon-sphere region and the AdS boundary has been shown to organize distinctive quasinormal-mode families; semiclassical analyses also identify long-lived AdS excitations whose damping is controlled by barrier penetration~\cite{Hashimoto2023,FestucciaLiu2009}. In the present calculation, we use the barrier height and the associated frequency ratio as diagnostics of the numerically obtained modes. These quantities characterize the leakage interpretation; they do not replace the global horizon-to-boundary eigenvalue condition derived in Sec.~\ref{sec:ads}.

\begin{figure}[tbp]
\centering
\includegraphics[width=\columnwidth]{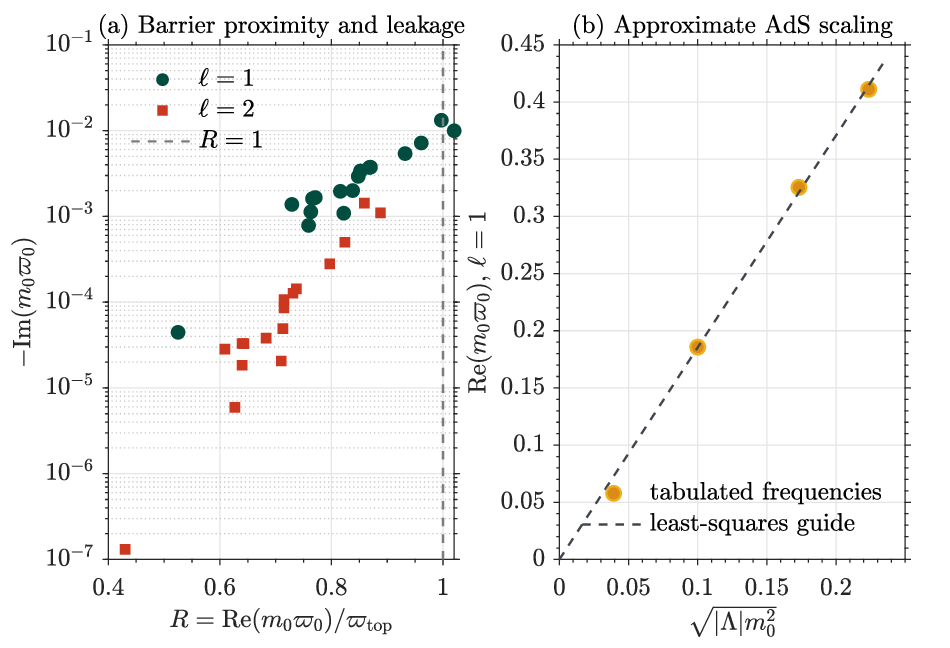}
\caption{AdS cavity diagnostics for the parameter sets in Table~\ref{tab:bar}. Panel (a) plots the fundamental damping rate $-Im({m}_{0}{\varpi }_{0})$ against the barrier-proximity ratio $R=\mathrm{Re}(m_0\varpi_0)/\varpi_{\rm top}$ for $\ell=1$ (circles) and $\ell=2$ (squares); the vertical dashed line marks $R=1$, where the real frequency reaches the barrier-top scale. The near-threshold cases, $\alpha =0.3$ for $\ell=1$ ($R=1.020$) and $\Lambda =-0.05$ for $\ell=1$ ($R=0.997$), are retained. Panel (b) shows $Re({m}_{0}{\varpi }_{0})$ for $\ell=1$ at $|\Lambda |{m}_{0}^{2}=0.01,0.03,0.05$; the dashed line is a least-squares guide proportional to $\sqrt{|\Lambda |{m}_{0}^{2}}$, not a universal scaling law. Frequencies and barrier data are from Tables~\ref{tab:ads} and \ref{tab:bar}.}
\label{fig:adswindow}
\end{figure}

The parameter dependence of these diagnostics is summarized in Fig.~\ref{fig:adswindow}. The ratio $R$ is used only as a comparative indicator of proximity to the barrier top; it is not an independent quantization condition and does not by itself determine the transmission probability.

The region between the photon sphere barrier and the AdS wall acts as a cavity, as in the qualitative trapping picture of Ref.~\cite{EslamPanah2026}. Waves with frequencies between the bottom of the cavity and the top of the barrier are partially trapped there and lose amplitude only by tunnelling through the barrier toward the horizon. In the tortoise coordinate this window is $\varpi_{\rm min}<\varpi<\varpi_{\rm top}$, where $\varpi_{\rm top}^2=V_{\rm max}$ is the height of the barrier and $\varpi_{\rm min}^2$ the value of $V$ at the minimum behind it. The corresponding values are given in Table~\ref{tab:bar}, together with the ratio ${\rm Re}\,\varpi_0/\varpi_{\rm top}$. For the reference set and $\ell=1$ the window is $0.249<\varpi <0.384$, and the exact fundamental frequency, ${\rm Re}\,\varpi_0=0.3254$, lies inside it. In Ref.~\cite{EslamPanah2026} the window was defined with the potential $U_\ell(r)$ of the wave equation written in the coordinate $r$, which is a different function from $V$ in Eq.~\eqref{eq:V}. For the reference set and $\ell=1$ that window is $0.255<\varpi<0.319$, so ${\rm Re}\,\varpi_0$ lies inside the $V$ window and just above the $U_\ell$ one. The turning points, and hence the window edges, depend on the form in which the wave equation is written. Such windows are therefore only rough guides, and, as noted in Ref.~\cite{EslamPanah2026}, the frequencies themselves follow from the boundary value problem of Sec.~\ref{sec:ads}.

\begin{table}[tbp]\centering\footnotesize\setlength{\tabcolsep}{3.5pt}
\caption{Bottom of the cavity $\varpi_{\rm min}$, top of the barrier $\varpi_{\rm top}=\sqrt{V_{\rm max}}$ and the ratio $R={\rm Re}\,\varpi_0/\varpi_{\rm top}$ for the fundamental AdS modes of Table~\ref{tab:ads}.}
\label{tab:bar}
\begin{tabular}{lcccccc}\hline\hline
 & \multicolumn{3}{c}{$\ell=1$} & \multicolumn{3}{c}{$\ell=2$}\\ \cline{2-4}\cline{5-7}
 & $\varpi_{\rm min}$ & $\varpi_{\rm top}$ & $R$ & $\varpi_{\rm min}$ & $\varpi_{\rm top}$ & $R$\\\hline
reference & 0.2487 & 0.3837 & 0.848 & 0.3489 & 0.6039 & 0.715\\
$\sigma =0$ & 0.2708 & 0.4155 & 0.852 & 0.3742 & 0.6498 & 0.715\\
$\sigma =0.06$ & 0.2033 & 0.3171 & 0.838 & 0.2955 & 0.5066 & 0.713\\
$\sigma =0.10$ & 0.1547 & 0.2448 & 0.822 & 0.2354 & 0.3983 & 0.710\\
$\alpha =0$ & 0.2570 & 0.4451 & 0.771 & 0.3572 & 0.6965 & 0.643\\
$\alpha =0.2$ & 0.2397 & 0.3278 & 0.932 & 0.3399 & 0.5189 & 0.797\\
$\alpha =0.3$ & 0.2296 & 0.2783 & 1.020 & 0.3299 & 0.4422 & 0.888\\
$q=0$ & 0.2487 & 0.3727 & 0.870 & 0.3488 & 0.5851 & 0.737\\
$q=0.6$ & 0.2488 & 0.4004 & 0.816 & 0.3490 & 0.6329 & 0.683\\
$q=0.8$ & 0.2488 & 0.4336 & 0.759 & 0.3490 & 0.6915 & 0.627\\
$\beta =0.05$ & 0.2451 & 0.3239 & 0.961 & 0.3455 & 0.5100 & 0.824\\
$\beta =0.4$ & 0.2521 & 0.4399 & 0.766 & 0.3521 & 0.6885 & 0.640\\
$\beta =0.6$ & 0.2547 & 0.4722 & 0.729 & 0.3546 & 0.7340 & 0.609\\
$\omega =-0.8$ & 0.2094 & 0.3677 & 0.763 & 0.2983 & 0.5818 & 0.640\\
$\omega =-0.5$ & 0.2646 & 0.3964 & 0.868 & 0.3677 & 0.6216 & 0.731\\
$\Lambda =-0.01$ & 0.1359 & 0.3540 & 0.525 & 0.1922 & 0.5622 & 0.430\\
$\Lambda =-0.05$ & 0.3246 & 0.4125 & 0.997 & 0.4551 & 0.6438 & 0.859\\
\hline\hline\end{tabular}\end{table}

\begin{figure}[tbp]
\centering
\includegraphics[width=0.7\columnwidth]{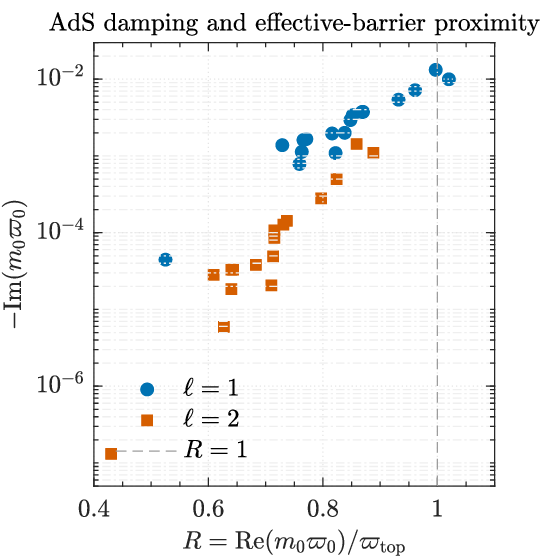}
\caption{Barrier-proximity correlation for the fundamental scalar AdS modes. The ordinate is the damping rate $-Im({m}_{0}{\varpi }_{0})$ on a logarithmic scale, and the abscissa is $R=\mathrm{Re}(m_0\varpi_0)/\varpi_{\rm top}$, where $\varpi_{\rm top}=\sqrt{V_{\rm max}}$ is the barrier-top frequency defined in Sec.~\ref{sec:cavity}. Circles and squares denote $\ell=1$ and $\ell=2$, respectively; the vertical dashed line marks $R=1$. Every point corresponds to a parameter set in Tables~\ref{tab:ads} and \ref{tab:bar}; no fitted damping law is imposed. The plot displays the five-decade spread in damping and the systematic increase in leakage as the real frequency approaches the barrier-top scale.}
\label{fig:adsdamp}
\end{figure}

With one exception, the fundamental $\ell=1$ and $\ell=2$ modes lie inside this window, as expected for trapped modes. The exception is the $\ell=1$ mode for $\alpha =0.3$ (ratio $1.02$), whose real part lies slightly above the barrier top; for $\Lambda =-0.05$ the $\ell=1$ mode lies just below the top (ratio $0.997$). The tunnelling picture suggests that the damping rate should be controlled by how close the frequency is to the top of the barrier. A wave far below the top has to cross a thick evanescent layer and is exponentially suppressed, while a wave close to the top leaks easily. Figure~\ref{fig:adsdamp} is consistent with this picture.

 When $|{\rm Im}\,\varpi_0|$ is plotted against ${\rm Re}\,\varpi_0/\varpi_{\rm top}$, the points for all parameter sets and both multipoles fall close to a single rising band, covering about five decades in the damping rate. The string cloud parameter lowers the barrier more than the trapped frequency, so the ratio grows (from $0.77$ at $\alpha =0$ to $1.02$ at $\alpha =0.3$ for $\ell=1$) and the modes become more strongly damped. The charge and the PFDM parameter raise the barrier and make the modes longer lived. The quintessence normalization $\sigma$ lowers the whole spectrum and the barrier by similar amounts and changes the ratio only a little. A larger $|\Lambda |$ pushes the wall inward and the trapped frequencies upward toward the barrier top, which is consistent with the strong dependence on $\Lambda$ found in the previous section. The two $\ell=1$ modes that lie at or just above the barrier top, for $\alpha =0.3$ and $\Lambda =-0.05$, are the most strongly damped of all $\ell=1$ fundamental modes, in line with the tunnelling picture.

The correlation is not perfect, since the thickness of the evanescent layer also depends on the shape of the potential and not only on its height, but it captures the main trends. It also suggests a reason for a qualitative difference between the two regimes. For $\Lambda =0$ the charge and the PFDM parameter increase the damping rate and the string cloud decreases it, and there the damping follows the curvature of the barrier top (the Lyapunov exponent). In AdS the trend is reversed for these three parameters (the quintessence normalization lowers the damping in both regimes), which is consistent with damping by tunnelling, where a higher barrier means slower leakage.

\section{Conclusions}
\label{sec:conc}

We have constructed and analysed the linear test-field spectra of charged Letelier black holes with a string-cloud contribution, logarithmic PFDM, and quintessence, the geometry of Ref.~\cite{EslamPanah2026}, considering both the $\Lambda =0$ and asymptotically AdS ($\Lambda <0$) regimes. Depending on the quintessence normalization, the $\Lambda =0$ geometry possesses either a cosmological horizon or an exterior extending to spatial infinity, whereas the $\Lambda <0$ geometry has a reflecting timelike AdS boundary. These distinct asymptotic structures define different spectral boundary-value problems and require different numerical treatments. The results are restricted to scalar and test-Maxwell perturbations on a prescribed background; they do not constitute a stability analysis of the fully coupled Einstein--Maxwell--matter system.

For $\Lambda =0$, the reference geometry has three positive horizons, ${r}_{-}=0.07518356{m}_{0}$, ${r}_{+}=1.70634028{m}_{0}$, and ${r}_{c}=43.95596099{m}_{0}$, with the static exterior region bounded by the event and cosmological horizons. The scalar and test-Maxwell potentials form single barriers in this region (see Fig.~\ref{fig:pot}), and their QNMs were obtained with sixth-order WKB and Pad\'e resummation. The independent characteristic evolutions provide a quantitative check: for the fundamental modes tabulated in Table~\ref{tab:td0}, the time-domain frequencies differ from the Pad\'e values by no more than $0.06\%$. The parameter scans in Fig.~\ref{fig:scan0} and Tables~\ref{tab:wkbs}--\ref{tab:wkbem} establish the direction and size of the shifts. Increasing $\sigma$ from $0$ to $0.10$ reduces the fundamental scalar $\ell=2$ frequency from $0.57464-0.11339i$ to $0.29771-0.05050i$; increasing $\alpha$ from $0$ to $0.3$ produces a comparable reduction, from $0.62335-0.12690i$ to $0.34707-0.05878i$. By contrast, increasing $q$ from $0$ to $0.8$ or $\beta$ from $0.05$ to $0.6$ raises both the oscillation frequency and the damping magnitude. These trends are not merely empirical: the photon-sphere angular frequency and Lyapunov exponent in Eq.~\eqref{eq:eik} reproduce the parameter dependence and approximate the $\ell=2$ WKB values at the percent level, as shown in Fig.~\ref{fig:eikonal}.

The sign of the PFDM parameter is physically consequential in the adopted convention. Equation~\eqref{eq:pfdm} gives $\rho_{\rm PFDM}=-\beta/(8\pi r^3)$, so the positive $\beta$ values in the principal scan correspond to a negative PFDM density and reduce the mass function near the photon sphere. The additional negative-$\beta$ calculations in Table~\ref{tab:negbeta} reverse this effect: the PFDM density becomes positive, the photon sphere and event horizon move outward, and the fundamental frequencies decrease. This sign-sensitive response is why the PFDM component is treated as an effective anisotropic source rather than identified without qualification with a conventional collisionless dark-matter halo. The horizon structure and the two asymptotic regimes are summarized in Fig.~\ref{fig:lapse}. The logarithmic time-domain amplitudes in Fig.~\ref{fig:td0} additionally show how the late-time response changes when the cosmological horizon is present or absent.

For $\Lambda <0$, the reflecting AdS boundary invalidates the outgoing-wave condition required by the single-barrier WKB formula. We instead solved the scalar boundary-value problem in ingoing Eddington--Finkelstein coordinates with a Chebyshev--Gauss--Lobatto discretization. The formulation imposes regularity at the future horizon and the normalizable (Dirichlet) branch at the AdS boundary. Its convergence is documented in Table~\ref{tab:conv}, and the Schwarzschild--AdS benchmarks in Table~\ref{tab:sads} agree with the established results. For the reference geometry the event horizon is ${r}_{+}=1.65563890{m}_{0}$. The spectral frequencies in Table~\ref{tab:ads} are independently checked against time-domain evolutions; the reference modes agree to four significant digits in their real parts and to within $3\times 10^{-4}$ in the relative imaginary parts, with the limitations for exceptionally long-lived modes at $\Lambda =-0.01$ stated in Sec.~\ref{sec:ads}. Figures~\ref{fig:adsspectrum} and \ref{fig:tdads} display, respectively, the parameter dependence of the fundamental spectrum and representative logarithmic-amplitude evolutions.

The AdS modes are best understood as quasi-bound cavity states between the photon-sphere barrier and the reflecting boundary. The barrier data in Table~\ref{tab:bar} and the correlations in Figs.~\ref{fig:adswindow} and \ref{fig:adsdamp} show that the damping rate generally increases as $R=\mathrm{Re}(m_0\varpi_0)/\varpi_{\rm top}$ approaches unity. The damping rates span approximately five orders of magnitude over the sampled backgrounds. The near-threshold cases $R=1.020$ ($\alpha =0.3$, $\ell=1$) and $R=0.997$ ($\Lambda =-0.05$, $\ell=1$) are important qualifications: the cavity ratio is a diagnostic rather than a strict necessary condition for a mode to be long-lived, and the barrier width and detailed potential shape also control transmission. The real parts of the $\ell=1$ fundamental frequencies at $\Lambda =-0.01,-0.03,-0.05$ are $0.18569$, $0.32538$, and $0.41112$, respectively, consistent over these three sampled points with an approximate $\sqrt{|\Lambda |}$ trend, as displayed in Fig.~\ref{fig:adswindow}. This finite data set does not establish a universal scaling law.

All computed frequencies have $Im\varpi <0$. This establishes the absence of growing modes only in the scalar and test-Maxwell sectors and parameter ranges actually examined. It does not prove mode stability of the coupled gravitational, electromagnetic, and matter perturbations, nor does it address nonlinear evolution or nonperturbative channels such as brane nucleation~\cite{XiaOng2025}. In particular, the electromagnetic calculation on the charged background is a test-field result; the coupled Einstein--Maxwell spectrum and a consistent perturbation model for each anisotropic source remain open problems. A complete treatment of those sectors, together with a systematic study of admissible AdS boundary conditions for propagating fields and a comparison with the thermodynamics of Refs.~\cite{EslamPanah2026,Ahmed2025,Toledo2019}, provides the natural next steps.

\section*{Data availability}
The Mathematica notebook used to obtain the results of this paper is available from the corresponding author on reasonable request.

\end{document}